\documentclass{aa}
\usepackage[varg]{txfonts}
\usepackage{graphicx}
\usepackage[normalem]{ulem}
\usepackage{natbib}

\def\cm{\,{\rm cm}\,}

\def\ccm{\,{\rm cm^{-3}}\,}

\def\degr{\hbox{$^\circ$}\,}
\def\farcm{\hbox{$.\mkern-4mu^\prime$}\,}
\def\farcs{\hbox{$.\!\!^{\prime\prime}$}\,}

\def\mJyb{\,{\rm mJy/beam\,area}\,}

\def\muG{\,$\mu$G\,}

\def\kpc{\,{\rm kpc}}

\def\ne{\,n_{\rm e}\,}
\def\radm{\,\rm{rad\,m^{-2}}\,}
\def\Bt{\,B_{\rm tot}}
\def\Br{\,B_{\rm ran}}
\def\Bra{\,B_{\rm ran,a}}
\def\Brp{\,B_{\rm ran,p}}
\def\Bo{\,B_{\rm ord}}
\def\Boa{\,B_{\rm ord,a}}
\def\Bop{\,B_{\rm ord,p}}
\def\alphan{\alpha_{\rm n}}
\def\pn{p_{\rm n}}
\def\DPn{DP_{\rm n}}
\def\RMi{RM_{\rm i}}
\def\MGpc{{\rm M_{\sun}}\,{\rm Gyr^{-1}}\,{\rm pc^{-2}}\,}
\def\SFRa{\Sigma_{\rm SFR}}
\def\SFRaj{\Sigma_{{\rm SFR},j}}
\def\HI{\ion{H}{I}\,}
\def\HII{\ion{H}{II}\,}

\begin{document}

\title{Radio polarization and magnetic field structure in M\,101
\thanks{Based on observations with the 100 m telescope of the MPIfR at Effelsberg}}

\author{E.M. Berkhuijsen\inst{1}
\and M. Urbanik\inst{2}
\and R. Beck\inst{1}
\and J.L. Han\inst{3}}

\titlerunning{Polarized emission from M\,101}
\authorrunning{E.M. Berkhuijsen et al.}

\institute{Max-Planck-Institut f\"ur Radioastronomie, Auf dem H\"ugel 69, 53121 Bonn, Germany \\
\email{eberkhuijsen@mpifr-bonn.mpg.de}
\and Astronomical Observatory, Jagiellonian University, ul. Orla 171, 30-244 Krak\'ow, Poland
\and National Astronomical Observatories, Chinese Academy of Sciences, A20 Datun Road, Beijing 100012, China}

\date{Received 7 September 2015 / Accepted 4 January 2016}

\abstract{We observed total and polarized radio continuum emission from the
spiral galaxy M\,101 at $\lambda\lambda$\,6.2\,cm and 11.1\,cm with the
Effelsberg telescope. The angular resolutions are $2\farcm5$ (=5.4\,kpc) and
$4\farcm4$ (=9.5\,kpc), respectively. We use these
data to study various emission components in M\,101 and properties of
the magnetic field. Separation of thermal and non-thermal emission shows that
the thermal emission is closely correlated with the spiral arms, while
the non-thermal emission is more smoothly distributed indicating diffusion
of cosmic ray electrons away from their places of origin. The radial
distribution of both emissions has a break near $R=16$\,kpc (=$7\farcm4$),
where it steepens to an exponential scale length of $L\simeq5$\,kpc, which is
about 2.5 times smaller than at $R<16$\,kpc. The distribution of the polarized
emission has a broad maximum near $R=12$\,kpc and beyond $R=16$\,kpc also
decreases with $L\simeq5$\,kpc. It seems that near $R=16$\,kpc a major
change in the structure of M\,101 takes place, which also affects the
distributions of the strength of the random and ordered magnetic field. Beyond
$R$=16\,kpc the radial scale length of both fields is about 20\,kpc, which
implies that they decrease to about 0.3\muG\, at $R$ = 70\,kpc, which is the
largest optical extent. The equipartition strength of the total field ranges
from nearly 10\muG at $R<2$\,kpc to 4\muG at $R=22-24$\,kpc. As the random field
dominates in M\,101 ($\Br$/$\Bo\simeq2.4$), wavelength-independent polarization
is the main polarization mechanism. We show that energetic events causing
$\HI$\,shells of mean diameter $< 625$\,pc could partly be responsible for this.
At radii $< \,24$\,kpc, the random magnetic field depends on the star formation
rate/area, $\SFRa$, with a power-law exponent of $b=0.28\pm0.02$. The ordered
magnetic field is generally aligned with the spiral arms with pitch angles that
are about $8\degr$ larger than those of $\HI$\,filaments.}

\keywords{Galaxies: individual: M\,101 -- galaxies: magnetic fields -- galaxies:
star formation -- polarization - radio continuum: galaxies -- radiation
mechanisms: non-thermal}

\maketitle


\section{Introduction}
\label{sect:intro}


\begin{table*}
\caption{Adopted parameters on M\,101}
\centering
\begin{tabular}{lll}
\hline\hline
Variable            &  Value                         &  Reference \\
\hline
Distance $D$ (Mpc)  &  7.4 ($1\arcmin=2.15$\,kpc)    & \citet{Kelson96} \\
Centre position \\
(RA,Dec)$_{2000}$   & $14^h\,03^m\,12\fs77, 54\degr\,20\arcmin\,54\farcs4$\
  & \citet{Israel75} \\
Position angle\, $PA$   & 38\degr                       & \citet{Kamphuis93} \\
Inclination\, $i^{\,a}$ & 30\degr\,(face-on $i=0\degr$) & \citet{Kamphuis93} \\
Radius in colour B: $R_{25}$   &  8\arcmin               & \citet{Mihos13} \\
Radius in colour
B: $R_{29.5}$ & 25\arcmin               & \citet{Mihos13} \\
Radius in \HI           & 27\arcmin               & \citet{Kamphuis93} \\
Hubble type             & SAB(rs)cd               & \citet{Vaucouleurs76} \\
\hline
\end{tabular}
\tablefoot{
\tablefoottext{a}{Kamphuis derived different inclination angles for radii
$R<7\arcmin$\, (27\degr) and $R>7\arcmin$\, (25\degr\, in SW and 40\degr\,
in NE). As our data extend to $R\sim 15\arcmin$, we adopted a mean value of
$i=30\degr$.}
}
\label{table:m101}
\end{table*}

The Pinwheel galaxy, M\,101 (NGC\,5457) is a nearby spiral galaxy seen
nearly face-on (see Table~\ref{table:m101}). It is an SAB(rs)cd galaxy
\citep{Vaucouleurs76} containing many \HII\,\,regions and several large
\HII\,\,complexes. Sensitive optical imaging of \citet{Mihos13} showed
that in blue light the bright parts of M\,101 have a radius of about
8\arcmin on the sky ($R_{25}=8\arcmin$), but that a weak optical disk
can be traced about three times further out ($R_{29.5}=25\arcmin$).
However, the galaxy is strongly lopsided, which may be due to past
encounters with one or more of the six companions forming the M\,101
group \citep[e.g.][]{Karachentsev14, Mihos13, Jog09, Waller97}. The
distance to M\,101 has been the subject of many optical studies. We have
adopted the Cepheid distance of $D = 7.4\pm0.6$\,Mpc derived by
\citet{Kelson96}, which is in good agreement with the compilation and
new measurements of \citet{Lee12}. Some basic parameters of M\,101
relevant to our work are listed in Table~\ref{table:m101}.

M\,101 has been observed at many wavelengths. High-resolution maps have
been presented in the emission lines of atomic hydrogen (\HI) \citep{Kamphuis93,
Braun95, Walter08}, CO(1-0) \citep{Kenney91, Helfer03} , and ionized hydrogen
(H$\alpha$) \citep{Scowen92, Hoopes01}, as well as of the emission in
far-ultraviolet (FUV) \citep{Waller97}, X-rays \citep{Kuntz03}, and mid-infrared
(mid-IR) \citep{Jarrett13}. These maps show a complicated structure of many
narrow, patchy spiral arms with large variations in pitch angle. The many
linear arm segments and the asymmetry of the large-scale structure are
attributed to a collision with the satellite galaxy NGC\,5474 \citep{Waller97,
Kamphuis93, Mihos12}. \citet{Kenney91} detected  a bar in the centre in CO,
which is also seen in H$\alpha$ and near-infrared (NIR), but density waves are
weak in M\,101 \citep{Kamphuis93}.

Little is known about the magnetic field in M\,101. The first radio
contiunum maps were presented by \citet{Israel75} who used aperture
synthesis at wavelengths $\lambda\lambda$\,49.2, 21.2, and 6\,cm showing
enhanced emission from spiral arms and \HII-region complexes.
\citet{Graeve90} carried out a multi-wavelength study of M\,101 at
$\lambda\lambda$\,11.1, 6.3, 2.8, and 1.2\,cm with the 100\,m telescope at
Effelsberg, leading to the first spectral index map of the galaxy. At
$\lambda$\,6.3\,cm they also obtained the first map of polarized emission
from M\,101, indicating the existence of an ordered, large-scale magnetic field
generally oriented along spiral arms. However, the sensitivity of these
data were insufficient for further analysis of the field properties.

We observed M\,101 with the Effelsberg telescope at $\lambda
\lambda$\,6.2 and 11.1\,cm with improved sensitivity in total power and
polarization. Our data allow a detailed study of the properties of the
magnetic field in M\,101 after separation of the thermal (free-free) and
non-thermal (synchrotron) components of the radio continuum emission.
In this paper, we study the strength and regularity of the magnetic field,
depolarization effects and their origin, the influence of star formation
on the strength of the random field, and the relationship between the
orientation of the ordered field and spiral arms.

The observations and reduction procedures are described in
Sect.~\ref{sect:obs}. In Sect.~\ref{subsect:maps} we present the resulting
maps and in Sect.~\ref{subsect:sep} we separate thermal/non-thermal emission
and derive radial scale lengths of the various emission components.
The discussion in Sect.~\ref{sect:disc} consists of several parts. Sect.~
\ref{subsect:Bstrength} shows the radial distribution of the magnetic field
strengths and the dependence of the random field on the star formation
rate (SFR); Sect.~\ref{subsect:RMDP} discusses Faraday rotation measures
and depolarization effects in M\,101; Sect.~\ref{subsect:pitch} describes
the large-scale structure of the ordered magnetic field, the alignment with
\HI arms, and a model to explain the alignment. Finally, we summarize our conclusions
in Sect.~\ref{sect:sum}.


\section{Observations and data reduction}
\label{sect:obs}

M\,101 was observed at the frequencies 2.7\,GHz ($\lambda$\,11.1\,cm) and
4.85\,GHz ($\lambda$\,6.2\,cm) with receiver systems in the 100\,m
Effelsberg telescope between July and December 1997. At these frequencies
the half-power beamwidths are 4\farcm4 and 2\farcm5, respectively. The
system parameters are listed in Table~\ref{table:syst}. The point sources
3C\,286 and 3C\,138 were observed for calibrations of flux density
and polarization angle. We adopted $S_\mathrm{11} = 5.8$\,Jy and
$S_\mathrm{6} = 3.8$\,Jy for 3C\,138, and $S_\mathrm{11} = 10.4$\,Jy and
$S_\mathrm{6} = 7.5$\,Jy for 3C\,286, respectively \citep{Ott94, Fernini97}.

We observed a large field of $51\arcmin\times 51\arcmin$\ at $\lambda$\,11.1\,cm
to enable proper base level determination. The field was centred on the galaxy
(see Table~\ref{table:m101}) and alternately scanned in RA and DEC. We used
a single horn, a scan speed of 60\arcmin\, per minute, and a scan separation
of 1.5\arcmin\, in DEC (or RA) between scans, which is about one-third of the
beamwidth, as needed for complete sampling of the emission. We obtained
14 coverages, half of which were scanned in RA and the other half in DEC.
Each coverage took about 40 minutes.

We carried out the data processing with the NOD2 package \citep{Haslam74}. After
removal of strong interference and adjustment of  base levels of individual
scans, final maps in Stokes $I$, $Q$, and $U$ were made with the baseline
optimizing procedure described by \citet{Emerson88}. After combining all
coverages, we reached noise levels of $\sigma_{\rm I}$ = 1.20\mJyb
and $\sigma_{\rm PI}$ = 0.54\mJyb for the maps of total intensity ($I$) and
polarized intensity ($PI=\sqrt{Q^2+U^2}$), respectively. Finally, the $PI$
map was corrected for positive noise bias \citep{Wardle74}.

We observed the same field at $\lambda$\,6.2\,cm  as at $\lambda$\,11.1\,cm,
using the dual-horn system. Because the beams of the two horns are separated
by 8\farcm12 in azimuth, the galaxy can only be scanned in azimuth. With a scan
speed of 60\arcmin\ per minute and a scan separation in elevation of 1\arcmin\,,
one coverage took about 51 minutes. In all, we obtained 20 coverages.

The dual-horn system is less sensitive to interference and weather
changes than the single-horn system because disturbances are largely
removed in the difference (i.e. time aligned and then subtracted) maps
of the two horns. During data processing with NOD2, we removed residual
interference from the difference $I$, $Q$, and $U$ maps, adjusted the
base level of each scan, restored the sky map from the difference maps
using the method by \citet{Emerson79}, and transformed the maps to
equatorial coordinates. Maps from all coverages were then combined to the
final $I$, $Q$, and $U$ maps using the NOD2 routine {\sc TURBOPLAIT}.
We reached a noise level of $\sigma_{\rm I} = 0.50$\mJyb
for the $I$ map and of $\sigma_{\rm PI} = 0.07$\mJyb for the $PI$ map,
which is nearly three times better in $I$ and more than ten times better
in $PI$ than was obtained by \citet{Graeve90}. Again, the $PI$ map was
corrected for positive noise bias \citep{Wardle74}. The estimated
error in the absolute flux-density scale is 5 percent and instrumental
polarization in the extended emission is negligible.


\begin{table}
\centering
\caption{System parameters}
\begin{tabular}{lcc}
\hline\hline
                    &   $\lambda$\,11.1\,cm   &    $\lambda$\,6.2\,cm    \\
\hline
Feed                      &   Single horn           &    Dual horn     \\
System Temperature (K)    &   45              &    30              \\
Centre Frequency (GHz)    &   2.7             &    4.85            \\
Bandwidth  (MHz)          &   40              &    300             \\
Half-power beamwidth      &   $4\farcm4$      &    $2\farcm5$      \\
$\sigma_{\rm I}$ (\mJyb)     & 1.20        &    0.50           \\
$\sigma_{\rm PI}$ (\mJyb)    & 0.54        &    0.07           \\
\hline
\end{tabular}
\label{table:syst}
\end{table}


\section{Results}
\label{sect:results}


\subsection{Total emission and polarized emission}
\label{subsect:maps}

\begin{figure}
\centering
\includegraphics[width=0.45\textwidth]{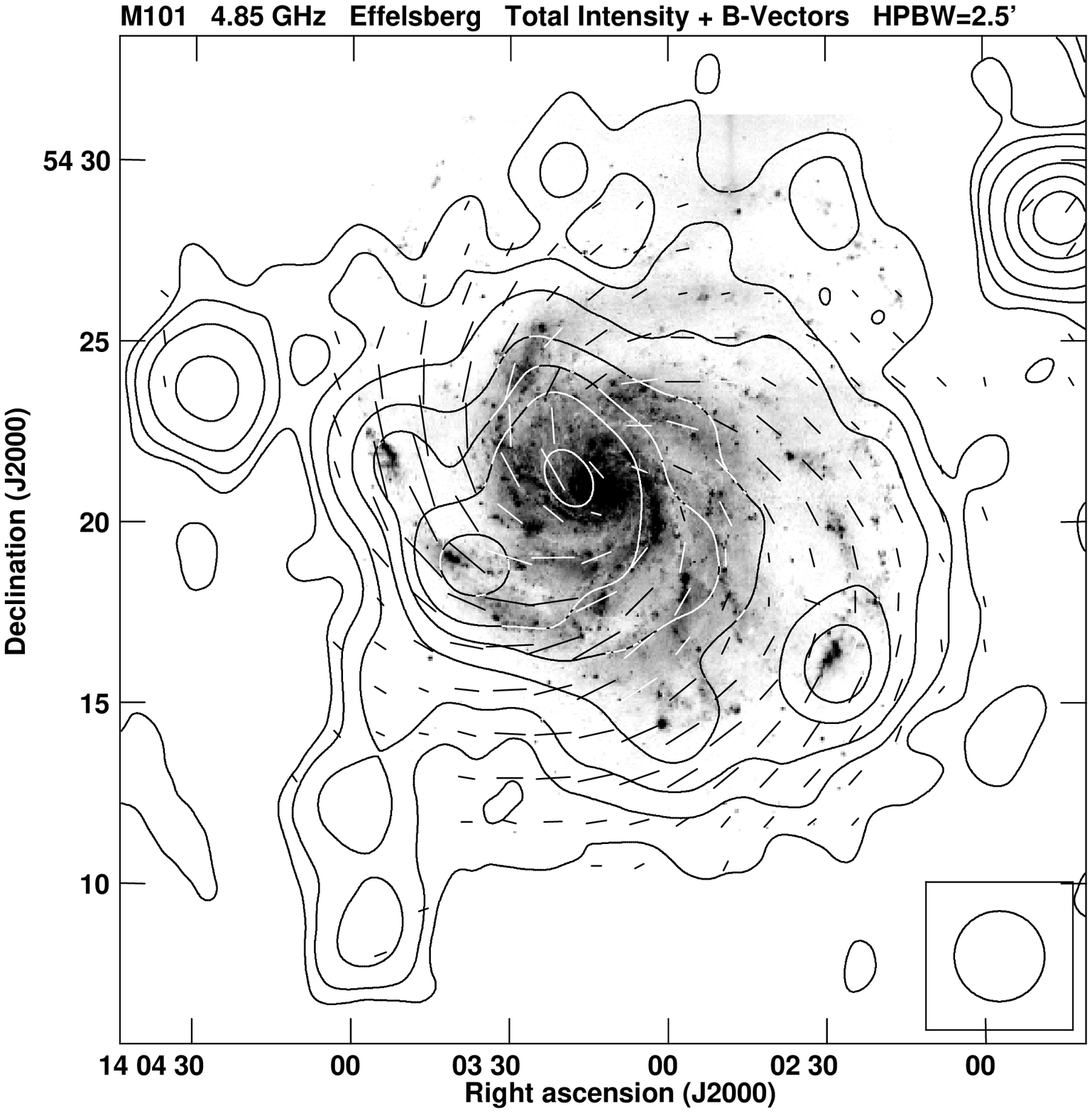}
\caption{Total  emission and apparent B-vectors of the polarized emission
(defined as E-vectors rotated by 90\degr) from M\,101 observed at $\lambda$
\,6.2\,cm overlaid on the optical image of \citet{Sandage61}. The contour
levels are 1, 2, 4, 8, 12, 16, and 24\,\mJyb. A vector of 1\arcmin\,
length corresponds to a polarized intensity of 1\mJyb. The noise levels are
0.5\,\mJyb in $I$ and 0.07\mJyb in $PI$. The beamwidth of $2\farcm5$ is shown
in the lower right corner. A square-root scale has been applied to the optical
 image to show low surface brightness structures more clearly.}
\label{fig:tp6}
\end{figure}

The distribution of the total radio emission from M\,101 at $\lambda$\,
6.2\,cm (Fig.~\ref{fig:tp6}) is asymmetric. The eastern half has a steep
brightness gradient towards the outside, while in the western half the
emission falls off more gradually beyond the western spiral arms. This
reflects the optical asymmetry in M\,101 with the western arms extending
to a considerably larger radius than the eastern arms. The maximum located
1\farcm5\, NE of the centre coincides with background source number 20 in
the list of \citet{Israel75}; the emission from the nucleus is much weaker.
Other brightness peaks coincide with large star-forming complexes in the
western arms as well as with two large complexes in the eastern arms and the
giant \HII\, region NGC\,5471 at RA=$14^h\,04^m\,28\fs6$, Dec=$54\degr\,
23\arcmin\,40\farcs3$. The south-eastern extension has no optical
counterpart; inspection of a larger field in the digitized sky survey (DSS)
and of the deep survey of \citet{Mihos13} did not show any optical emission
along this feature. It consists of several background sources unrelated to
M\,101. Checking the catalogue of faint images of the radio sky at twenty cm
(FIRST), we found two compact sources coinciding with the upper maximum in the
extension and three sources with the lower maximum. The strong source in
the north-west on the edge of the field also is a background source.

\begin{figure}
\centering
\includegraphics[width=0.45\textwidth]{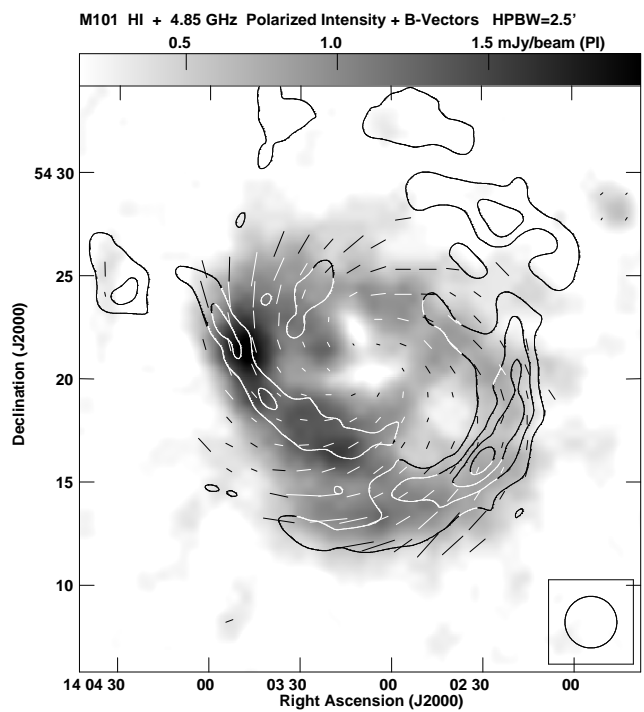}
\caption{Greyscale plot of the observed intensity of polarized emission
from M\,101 at $\lambda$\,6.2\,cm and apparent B-vectors (E+90\degr, not
corrected for Faraday rotation) with length proportional to the degree of
polarization. A vector of 1\arcmin\, length corresponds to 20\%. The noise
level in PI is 0.07\mJyb. Contours show the brightness distribution of \HI\,
of \citet{Braun95}. The contour levels of column density are (10, 15, 20, and
25)\,$10^{20}\,\cm^{-2}$. The beamwidth of $2\farcm5$ is shown in the lower
right corner.}
\label{fig:pipc6}
\end{figure}

\begin{figure}
\centering
\includegraphics[width=0.45\textwidth]{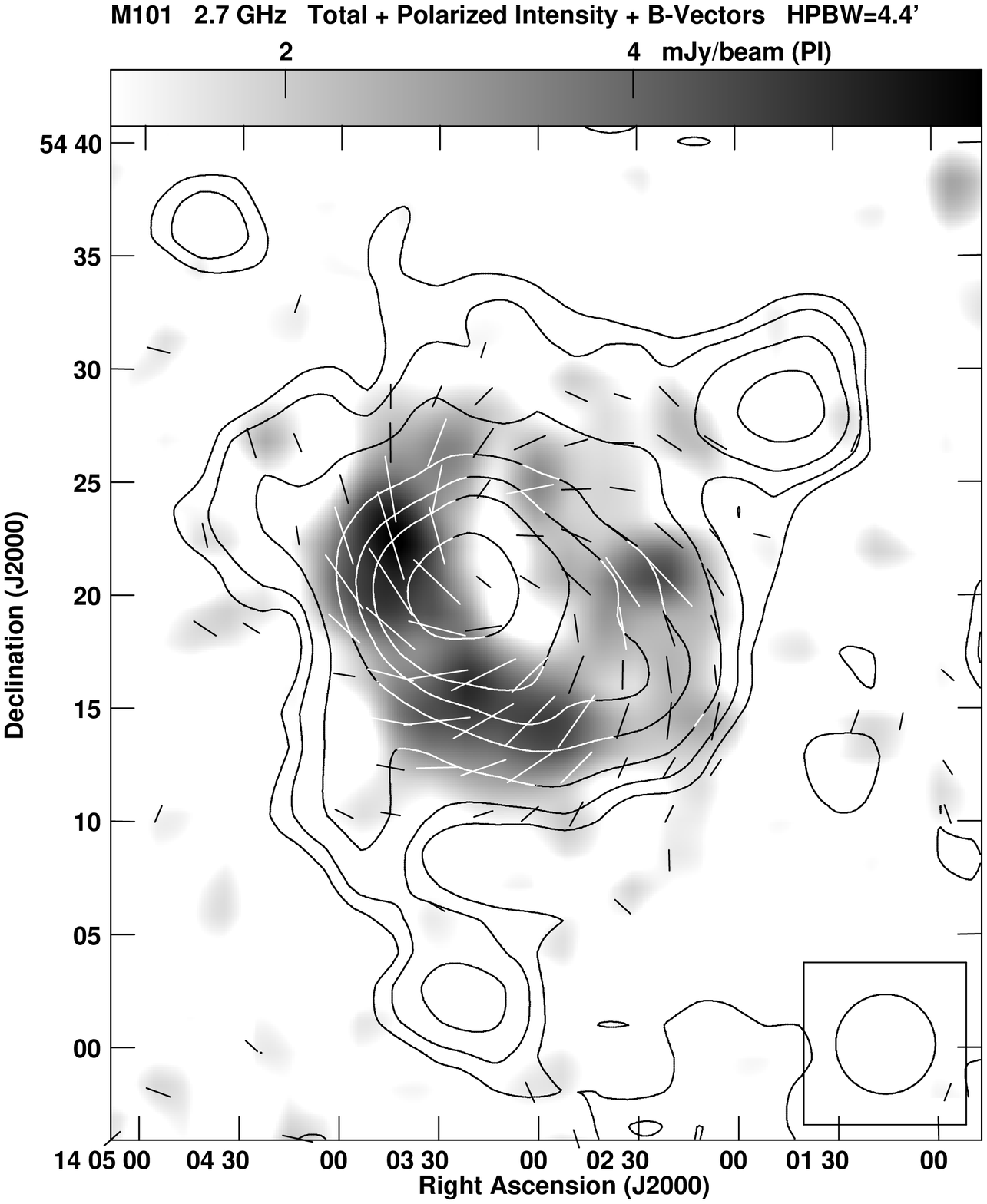}
\caption{Distribution of the total emission and apparent B-vectors
(E+90\degr, not corrected for Faraday rotation) of polarized emission from
M\,101 observed at $\lambda$\,11.1\,cm, overlaid on a greyscale image of
the polarized intensity. Contour levels are 3, 6, 12, 24, 36, 48, and 72\,\mJyb,
a vector of 1\arcmin\, length corresponds to a polarized intensity of
1.5\mJyb. The noise levels are 1.2\mJyb in $I$ and 0.54\mJyb in $PI$. The
beamwidth of $4\farcm4$ is shown in the lower right corner.}
\label{fig:tp11}
\end{figure}

The asymmetry in total emission and the extension towards the south are also
visible in the map at $\lambda$\,11.1\,cm (Fig.~\ref{fig:tp11}). The extended
maximum near the centre is clearly displaced from the nucleus because of the
background source mentioned before. In the western disk, the radio contours
show some emission enhancement at the position of a large star-forming
complex in the spiral arm; the emission is also enhanced on NGC\,5471.

The distribution of polarized emission from M\,101 at $\lambda$\,6.2\,cm
shows the same east-west asymmetry as the total emission
(Fig.~\ref{fig:pipc6}). The brightest peak is located at 5\arcmin\, east
of the optical centre on the inside of the outer eastern arm. The innermost
disk is depolarized by various effects (see Sect.~\ref{subsubsect:depol}).
The size of the southern part of this minimum corresponds to the area
below the central bar and innermost spiral arms seen in H$\alpha$
\citep{Scowen92} and CO \citep{Kenney91}, while the upper part of the minimum
is on a crossing of several thin arms just north of the innermost arm
Another depression in the polarized intensity occurs
about 5\arcmin\, south-west of the centre. It does not correspond to any
particular optical or H$\alpha$ feature,
but coincides with an extended minimum in the \HI\, map of \citet{Braun95}
between two major spiral arms.

The apparent polarization B-vectors (defined as observed E-vectors rotated
by 90\degr) at $\lambda$\,6.2\,cm form a very regular spiral pattern
(Figs.~\ref{fig:tp6} and \ref{fig:pipc6}). Despite the moderate resolution,
the apparent magnetic field orientations follow the optical spiral arms.
The same magnetic pattern is observed at $\lambda$\,11.1\,cm
(Fig.~\ref{fig:tp11}). The similar orientations of the vectors suggest that
Faraday rotation between these frequencies is small (see Sect.~\ref{subsubsect:depol}).


\subsection{Thermal and non-thermal emission}
\label{subsect:sep}

Before further analysing our data, we subtracted four unrelated point
sources from the total power maps at $\lambda\lambda$\,6.2 and 11.1\,cm.
We then smoothed the $\lambda$\,6.2\,cm maps in $I$ and $PI$ to a beamwidth
of $2\farcm7$ and those at 11.1\,cm to $5\farcm0$, which improved the
sensitivities at $\lambda$\,6.2\,cm to $\sigma_{\rm I} (\sigma_{\rm PI})$
= 0.460 (0.065)\mJyb and at $\lambda$\,11.1\,cm to $\sigma_{\rm I}\,
(\sigma_{\rm PI})$ = 1.05 (0.47)\mJyb.

For the separation of thermal and non-thermal components of the total
emission we need a map of the total spectral index $\alpha$ and the
non-thermal spectral index $\alphan$
\footnote{We use the convention $S\propto \nu^{-\alpha}$}.
\citet{Graeve90} derived a spectral index map between
$\lambda\lambda$\,49.2\,cm and 2.8\,cm at $1\farcm5$ resolution (see their
figure 5a). After smoothing the $\lambda\lambda$\,49.2\,cm and 2.8\,cm maps
to the resolutions of $2\farcm7$ and $5\farcm0$, which considerably reduced
the noise, we calculated maps of total spectral index at our resolutions for
all points above the noise level in both maps. The spectral index varies
from about 0.6 in the inner part to 0.9 or 1.0 at large radii. The large
difference in $\lambda$ between the maps and the low noise yield errors in the
$\alpha$ map of <\,0.02 within 7 arcmin from the centre, which slowly increase
to <\,0.1 further out.

\citet{Graeve90} determined $\alpha$ and $\alphan$ with the method described
by \citet{Klein84}, using the integrated flux densities for $R < 14'$ at ten
frequencies. They found $\alpha = 0.72\pm 0.04$ and $\alphan = 0.92\pm 0.18$.
Furthermore, \citet{Graeve90} observed that $\alpha$ becomes about 0.9 in
the outer parts of M\,101 where all the emission is non-thermal, and they
found that after subtraction of the bright \HII regions $\alpha$ also becomes
about 0.9 in the inner parts. So $\alphan$ must be close to 0.9. Following
\citet{Graeve90}, we integrated our $\lambda\lambda$\,6.2 and 11.1\,cm maps
over the area $R<14\arcmin$\,, yielding $S_{\rm 6} = 310\pm 20$\,mJy and
$S_{\rm 11} = 480\pm 30$\,mJy. These values are less than $8\%$ lower than
those listed by \citet{Graeve90} but agree within errors. Therefore we
adopted the value of $\alphan = 0.92\pm 0.10$ for our study.

For the separation of thermal/non-thermal emission only pixels in the
spectral index map with realistic values of $\alpha$ were used. If
$\alpha \le 0.1$ the emission is fully thermal and fully non-thermal if
$\alpha \ge \alphan$; elsewhere the thermal fraction is calculated. The resulting
thermal emission is then subtracted from the total emission to obtain the
non-thermal emission. In Sect.~\ref{subsubsect:alphan} we discuss how thermal
and non-thermal emission depend on the uncertainty of 0.1 in $\alphan$.

\begin{figure}
\centering
\includegraphics[width=0.45\textwidth]{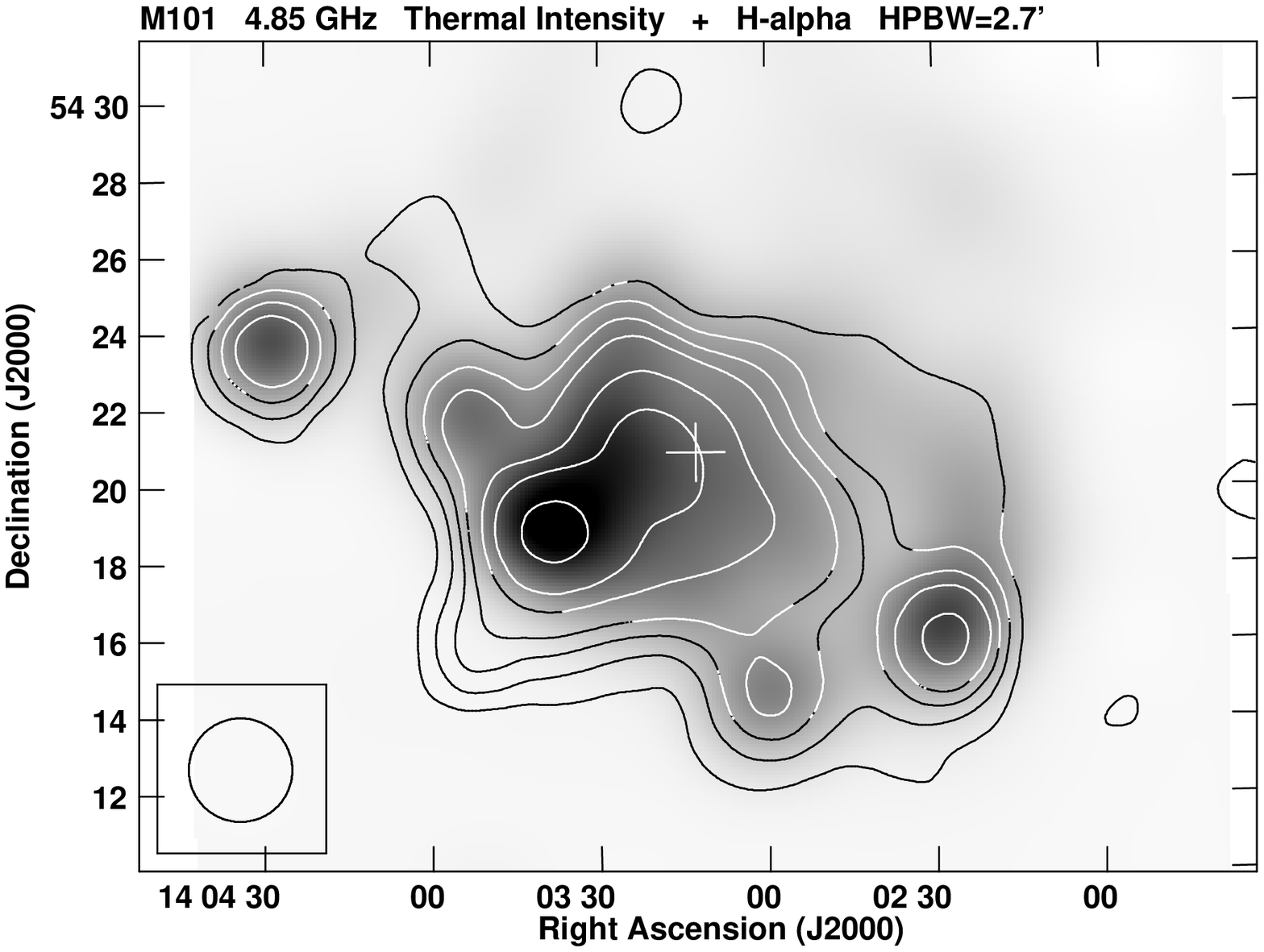}
\caption{Distribution of the thermal radio emission from M\,101 at
$\lambda$\,6.2\,cm overlaid on a greyscale plot of the H$\alpha$ emission
of \citet{Hoopes01} smoothed to the same beamwidth of $2\farcm7$ (shown in the
lower left corner). Contour levels are (1, 2, 3, 4, 6, 8, and 12)\,\,$\times$
1.5\,\mJyb. The noise level is about 0.5\,\mJyb. The white plus shows the
position of the optical centre. The strong source near the eastern
border of the map is the \HII-region complex NGC\,5471.}
\label{fig:therm}
\end{figure}

In Fig.~\ref{fig:therm} we compare the distribution of thermal emission at
$\lambda$\,6.2\,cm with that of the H$\alpha$ emission \citep{Hoopes01}
smoothed to the same beam size. Maxima in the radio thermal emission from
M\,101 agree well with those in the H$\alpha$ emission.

\begin{figure}
\centering
\includegraphics[width=0.45\textwidth]{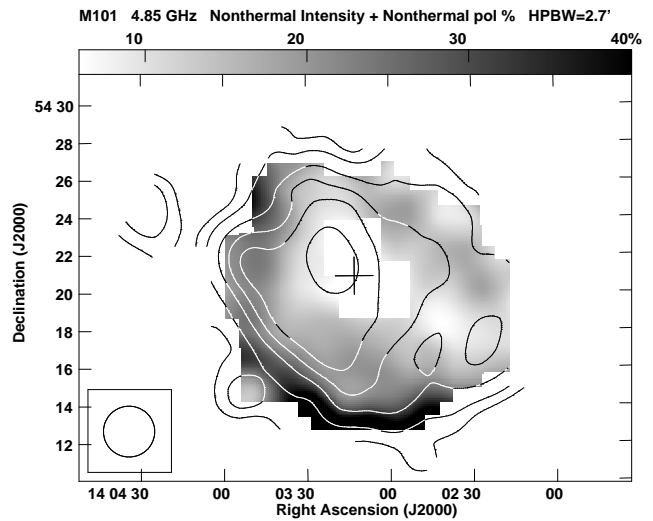}
\caption{Distribution of the non-thermal emission from M\,101 at
$\lambda$\,6.2\,cm (contours) superimposed onto the non-thermal degree of
polarization (greyscale). Contour levels are 1, 2, 4, 6, 8, and 12\,\mJyb.
The centre position is indicated with a plus. The maximum in the emission
NE of the centre is due to the background source number 20 in the list of
\citet{Israel75}. The noise level in the non-thermal intensity is about
0.5\mJyb. In the centre region, strong depolarization causes very low
non-thermal polarization degrees. The beamwidth is $2\farcm7$.}
\label{fig:nthpn}
\end{figure}

In Fig.~\ref{fig:nthpn} we show the distribution of the non-thermal emission
from M\,101, $NTH$, at $\lambda$\,6.2\,cm (contours) superimposed onto the
degree of non-thermal polarization $\pn = PI/NTH$ (greyscale). The $NTH$ has a
larger extent than the thermal emission, especially to the north. The
strong emission 1\farcm5 NE of the centre is from the background source
number 20 listed by \citet{Israel75}; the emission from the nucleus itself is
very weak. The $NTH$ is slightly enhanced on the brightest spiral arms and on
the star formation complex in the south-west, which is visible in
Fig.~\ref{fig:therm}. The values of  $\pn$ gradually increase from the
centre outwards and degrees of more than 40\% are reached in the south.
On the star formation complex in the south-west, $\pn$ has a minimum of
$<\,10\%$.

The integrated flux densities of the thermal ($TH$), $NTH$, and polarized
($PI$) emission at $\lambda\,6.2$\,cm are listed in Table~\ref{table:sint},
together with the average thermal fraction $f_{\mathrm{th}}=TH/I$ and the mean
value of $\pn=PI/NTH$. About 50\% of the $TH$ comes from the five giant \HII
region complexes observed by \citet{Israel75}. The flux density of $TH$ and
$f_{\rm th}$ may be overestimated by 20--25\% because we used a constant value
of $\alphan$, which is too large for star-forming regions
\citep{Tabatabaei07b, Tabatabaei13a}. In this case, $NTH$ ($\pn$) is
underestimated (overestimated) by nearly 20\%. For further interpretation, a
more realistic separation of thermal/non-thermal emission is required, i.e. by
determining the thermal emission from extinction-corrected H$\alpha$ data,
which does not need the assumption of a   constant value of $\alphan$ throughout
the galaxy \citep{Tabatabaei07b}.


\begin{table}
\centering
\caption{Integrated flux densities of M\,101 for $R < 14\arcmin$\,(=30\,kpc)}
\begin{tabular}{llc}
\hline\hline
Component       &  $\lambda$\,6.2\,cm  & Systematic error \\
\hline
$I$ (mJy)           &  $310\pm20$      &      ---       \\
$TH$ (mJy)          &  $140\pm15$      &    $\pm 30$    \\
$f_{\rm th}$        &  $0.45\pm0.06$   &    $\pm 0.1$   \\
$NTH$ (mJy)         &  $170\pm15$      &    $\pm 30$    \\
$PI$  (mJy)         &  $ 28\pm 3$      &      ---       \\
$\pn$               &  $0.16\pm0.02$   &    $\pm 0.03$  \\
\hline
\end{tabular}
\tablefoot{
Errors in column 2 are statistical errors; those in column 3 arise
from an uncertainty of 0.10 in $\alphan$ (see Sect.~\ref{subsubsect:alphan}).
}
\label{table:sint}
\end{table}

\begin{figure}
\centering
\includegraphics[angle=270,width=0.45\textwidth]{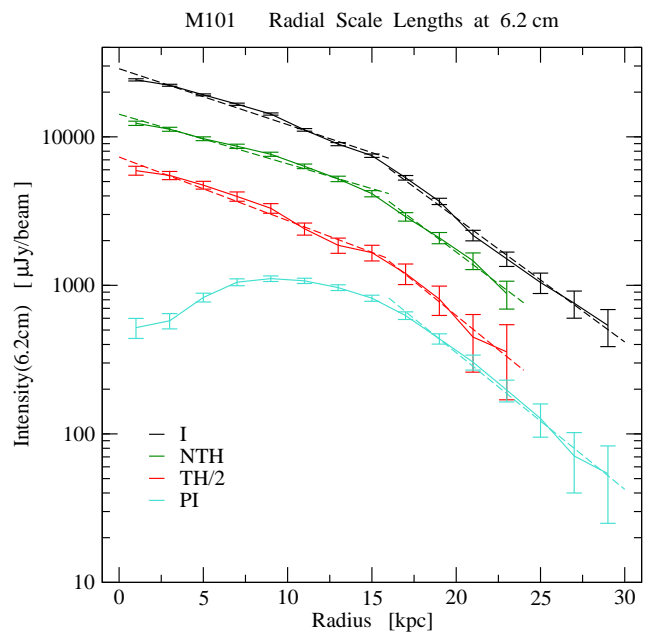}
\caption{Average intensity in 2\,kpc-wide rings in the plane of M\,101 of
$I$ and the emission components $NTH$, $TH$ and $PI$ plotted against
galacto-centric radius. For clarity, intensities of $TH$ are halved. Errors
are standard deviations. Note the change in the slopes near $R$=16\,kpc.
Dashed lines show the fits giving the exponential radial scale lengths listed
in Table~\ref{table:Rscale}.}
\label{fig:radial}
\end{figure}

The radial distributions of $I$ and the emission components in M\,101
at $\lambda$\,6.2\,cm are shown in Fig.~\ref{fig:radial}. The deep central
minimum in $PI$ is clearly visible, but $NTH$ and $TH$ similarly decrease
with increasing radius. We do not show $TH$ and $NTH$ points for $R > 24$\,kpc
because at these large radii the $TH$ and $NTH$ maps are no
longer complete, which make the radial averages unreliable. In each of the
curves a break is visible near $R=16$\,kpc. Therefore, we separately determined
exponential radial scale lengths $L$ for the intervals $R=0-16$\,kpc,
$R=16-24$\,kpc and $R=16-30$\,kpc (for $I$ and $PI$) by fitting the
intensities, weighted by their errors, to $I(R)=a \cdot \exp{(-R/L)}$. For
$PI$ only $L$ at large $R$ could be determined. The resulting scale lengths
are given in Table~\ref{table:Rscale}. At $R=0-16$\,kpc, $NTH$ decreases more
slowly ($L=13.0\pm1.4$\,kpc) than $TH$ ($L=10.2\pm1.0$\,kpc), as is expected
if cosmic ray electrons diffuse away from their birth places in star-forming
regions. However, beyond $R=16$\,kpc all three components have the same
radial scale length of $L\simeq 5$\,kpc, suggesting that in the outer
disk the cosmic ray electrons escape into the halo of M\,101.

\citet{Mihos13} found a change in the radial scale length of the optical
surface brightness at $R=7\arcmin\,-9\arcmin$, which is the same radius as
the break in the radio profiles. This position near $R=16$\,kpc (=$7\farcm4$)
corresponds to the radius where the inclination angle changes and the
\HI gas starts deviating from differential rotation \citep{Kamphuis93}.
Beyond $R=7\arcmin$\, the gas starts flaring with velocity components
perpendicular to the midplane of M\,101.

The change in scale length near $R$ = 16\,kpc, which is seen in the
distributions of thermal and non-thermal radio emission and optical surface
brightness, is accompanied by a change in the velocity structure near the same
radius. Taken together, these phenomena indicate a major change in the structure
of M\,101 near $R$=16\,kpc.

A break in the scale length of the radio continuum emission near the radius
where the star formation vanishes has also been found in M\,33
\citep{Tabatabaei07c}, M\,51 \citep{Mulcahy14} and IC\,342 \citep{Beck15}. The
IR emission from M\,33 also shows a break at this radius. Hence, a break in the
radial scale length of emission components near the radius where the star
formation comes to an end may be a general phenomenon in galaxies.


\begin{figure}
\centering
\includegraphics[angle=270,width=0.45\textwidth]{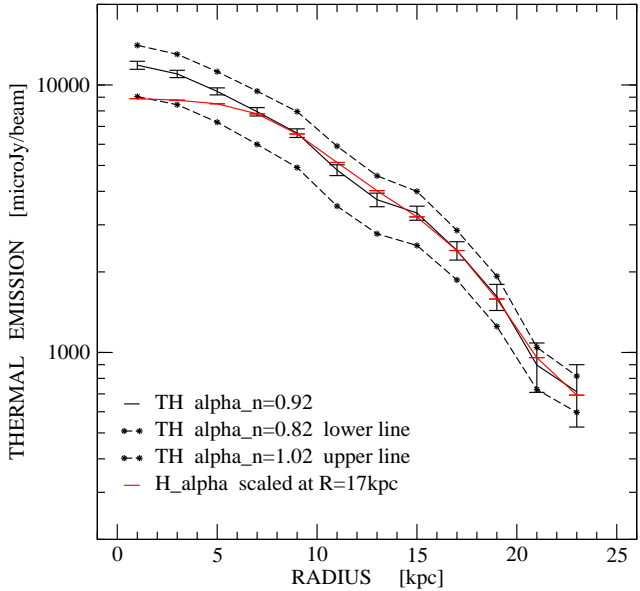}
\caption{Average intensity in 2\,kpc-wide radial rings in the plane of
M\,101 of $TH$ derived with $\alphan$ = 0.92 (solid line) and  with
$\alphan$= 0.82 and 1.02 (dotted lines), plotted against galacto-centric
radius. The errors on the dotted lines are the same as on the solid line,
but are not shown for clarity. The red line shows the radial distribution
of the H$\alpha$ emission \citep{Hoopes01}, scaled to $TH$ derived with
$\alphan$ = 0.92 at the ring $R = 16-18$\,kpc. Note the close corresponcence
between the two distributions at $R > 6$\,kpc. The discrepancy in the central
part may be due to extinction in H$\alpha$ and a possible overestimate
of $TH$ on \HII regions.}
\label{fig:errorTH}
\end{figure}


\subsubsection{The effect of an error in $\alphan$ on TH and NTH}
\label{subsubsect:alphan}

We  repeated the thermal/non-thermal separation for $\alphan$ = 0.82 and
$\alphan$ = 1.02 to investigate how sensitive $TH$ and $NTH$ are to the error
in $\alphan$. Figure~\ref{fig:errorTH} shows
the radial variation of the thermal emission for these cases and for
$\alphan$ = 0.92. The difference from our standard case  is typically
20\%, hence, the error in $\alphan$ causes a systematic error of 20\% in
$TH$ and a similar error in $NTH$. As the non-thermal degree of polarization
is $\pn = PI/NTH$,  $\pn$ also has a 20\% systematic error (see Table~
\ref{table:sint}). The resulting systematic errors in the scale lengths are
given in Table~\ref{table:Rscale}.

In Fig.~\ref{fig:errorTH} we also show the radial profile of the H$\alpha$
emission observed by \citet{Hoopes01}, scaled to $TH$ for $\alphan = 0.92$
at $R = 16-18$\,kpc. Apart from the inner 6\,kpc the profiles are almost
identical. The discrepancy near the centre is due to the combination of
extinction in H$\alpha$ and a possible overestimate of $TH$ on the many
\HII\  regions in this area in M\,101. This comparison, and the overlay in
Fig.~\ref{fig:therm}, show that our thermal/non-thermal separation yields a
good estimate of the distribution of the thermal emission in the galaxy.


\section{Discussion}
\label{sect:disc}

We now employ the non-thermal and polarized emission components derived in
the foregoing sections for an analysis of various properties of the
magnetic field in M\,101. We show how magnetic field strengths decrease
with increasing distance to the centre and how the random magnetic field
depends on the star formation rate per unit area, $\SFRa$. We discuss Faraday
rotation measures and depolarization effects, and look at the large-scale
structure of the ordered field.

\subsection{Magnetic field strengths and star formation rate}
\label{subsect:Bstrength}

\subsubsection{Radial distribution of magnetic field strengths}
\label{subsubsect:radialB}

\begin{figure}
\centering
\includegraphics[width=0.445\textwidth,angle=270]{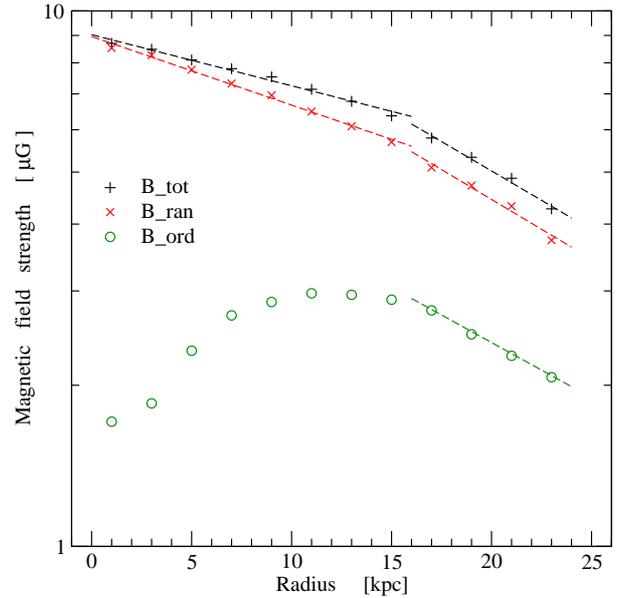}
\caption{Variation with galacto-centric radius of the equipartition magnetic
field strengths $\Bt$, $\Br$, and $\Bo$ averaged in 2\,kpc-wide rings in the
plane of M\,101. There is a change in slope near $R$=16\,kpc. Dashed lines
represent the exponential fits yielding the radial scale lengths given in
Table~\ref{table:Rscale}}.
\label{fig:Brad}
\end{figure}

From the radial variations of the surface brightnesses of $NTH$ and $PI$
at $\lambda$\,6.2\,cm presented in Fig.~\ref{fig:radial}, we calculated the
mean equipartition strengths of the total ($\Bt$), ordered ($\Bo$),
\footnote{{\em Ordered}\ magnetic fields as traced by linearly polarized
emission can be either {\em regular}\ fields, preserving their direction
over large scales (leading to both polarized emission and rotation
measure), or {\em anisotropic random}\, fields with multiple random field
reversals within the telescope beam, caused by shear and/or compression of
isotropic random fields (leading to polarized emission but not to rotation
measure). To observationally distinguish between these fundamentally
different types of magnetic field, additional Faraday rotation data is needed.}
and random ($\Br$) magnetic field using the code BFIELD of M. Krause
based on equation (3) of \citet{Beck05}. The code also requires the non-thermal
spectral index $\alphan$, the non-thermal degree of polarization
$\pn$, the line of sight $L_{\rm nth}$ through the emitting medium, and the
ratio of the energy densities of protons and electrons $K$, here taken
as 100. We used $\alphan=0.92$ (Sect.~\ref{subsect:sep}) and a scale
height of the non-thermal emission of 1\,kpc, leading to $L_{\rm nth}$=
$2/cos(i)=2.3$\,kpc. Fig.~\ref{fig:Brad} shows the radial distributions
of $\Bt$, $\Br$, and $\Bo$ in 2\kpc-wide rings around the centre for
$R < 24$\,kpc.
\footnote{As field strength scales with the power $1/(3+\alphan)$ of $K$,
$L_{\rm nth}$, and $NTH$, errors in these quantities and observational errors
in $NTH$ have little effect on the derived field strengths. The uncertainty of
0.1 in $\alphan = 0.92$ leads to less than 2\% changes in $\Bt$ and $\Br$ at
$R < 16$\,kpc and less than 5\% errors at larger radii. Only the systematic
error in $\Bo$ is about 17\% due to the systematic error in $\pn$
(see Sect.~\ref{subsubsect:alphan}).} The total field strength is nearly
10\muG near the centre and drops to about 4\muG in the ring $R=22-24$\,kpc.
The mean field strengths in the area $R<24$\,kpc are $\Bt=6.4$\muG,
$\Br=5.9$\muG and $\Bo=2.5$\muG. With $\Br$/$\Bo=2.4$, the magnetic field
in M\,101 is highly random like in, for example IC\,342 \citep{Beck15}.

In Fig.~\ref{fig:Brad} the break in the slope of the curves near $R=16$\,kpc
is very clear. Like in Sect.~\ref{subsect:sep}, we calculated the
exponential radial scale lengths for the two intervals $R < 16$\,kpc and
$R = 16-24$\,kpc. Table~\ref{table:Rscale} shows that the magnetic
fields have very long scale lengths of $34-45\kpc$\, at $R < 16$\,kpc and
about 20\,kpc at larger radii. In the inner region, $\Bo$ is low due to
the depolarization; therefore, the scale length of $\Bt$ is significantly
larger than that of $\Br$ ($(\Bt)^2 = (\Br)^2 + (\Bo)^2$). In the outer region,
$\Br$ and $\Bo$ have the same scale length. If this scale length remains the
same out to the radius of the maximal observed optical extent of
$R\,\simeq\,70$\,kpc\, \citep{vanDokkum14} and of the \HI\, gas of the
extension in the southwest of $R\,=\,90$\,pc \citep{Mihos12}, the field
strengths will have dropped to about 0.3\muG and 0.2\muG, respectively.
Hence, the intragroup magnetic field strength is probably smaller than 0.3\muG,
which is similar to the value estimated for a local group of irregular dwarf
galaxies \citep{Chyzy11}.


\begin{table}
\caption{Exponential radial scale lengths $L$ [kpc] of surface brightness
at $\lambda\,6.2\,$cm and magnetic field strength. Errors are statistical
errors. The numbers immediately below $NTH, TH$, and the field strengths are
systematic errors in case $\alphan$ = 1.02 (first one) or 0.82 (second
one), respectively.   The ratio between the scale lengths of $\Bt$ and $NTH$
at $R=16-24$\,kpc is $(3+\alphan)$, which is expected if $\pn$ is constant.
Because $\pn$ increases at $R = 0-16$\,kpc, the ratio between the scale lengths
of $\Bt$ and $NTH$ is less than ($3+\alphan$).}
\begin{tabular} {lccc}
\hline\hline
          & $R=0-16$\,kpc   & $R=16-24$\.kpc  & $R=16-30$\,kpc     \\
\hline
$I$       &  $11.5\pm1.0$   &  $ 4.7\pm0.5$   &  $ 5.2\pm0.3$      \\
$NTH$     &  $13.0\pm1.4$   &  $ 5.1\pm0.7$   &      -----         \\
          &  $    +0.2-0.5$ &  $    +0.2-0.2$ &      -----         \\
$PI$      &      -----      &  $ 5.1\pm0.2$   &  $ 4.7\pm0.3$      \\
$TH$      &  $10.2\pm1.0$   &  $ 4.7\pm0.8$   &      -----         \\
          &  $    +0.1-0.3$ &  $    +0.4-0.1$ &      -----         \\
          &                 &                 &                    \\
$\Bt$     &  $45.5\pm3.6$   &  $19.8\pm2.9$   &      -----         \\
          &  $    +2.3-3.3$ &  $    +2.0-1.5$ &      -----         \\
$\Br$     &  $33.9\pm1.9$   &  $19.5\pm3.6$   &      -----         \\
          &  $    -1.2+0.2$ &  $    +2.8-1.8$ &      -----         \\
$\Bo$     &      -----      &  $21.1\pm0.8$   &      -----         \\
          &      -----      &  $    -1.7+0.6$ &      -----         \\
\hline
\end{tabular}
\label{table:Rscale}
\end{table}


\subsubsection{Dependence of magnetic field strength on star formation rate}
\label{subsubsect:BSFR}

Since supernova explosions, SNRs, and stellar winds are the principal actors
stirring up the ISM, and hence producing random magnetic fields, a
relationship between the random magnetic field $\Br$ and the mean star
formation rate per unit area, $\SFRa$, is expected. This has indeed been
found for the galaxies NGC\,4254 \citep{Chyzy08} and NGC\,6946
\citep{Tabatabaei13b} as well as for the global values of a sample
of nearby galaxies (e.g. \citet{Heesen14}). Below we show that a relationship
also exists in M\,101.

\begin{figure}
\centering
\includegraphics[width=0.45\textwidth,angle=270]{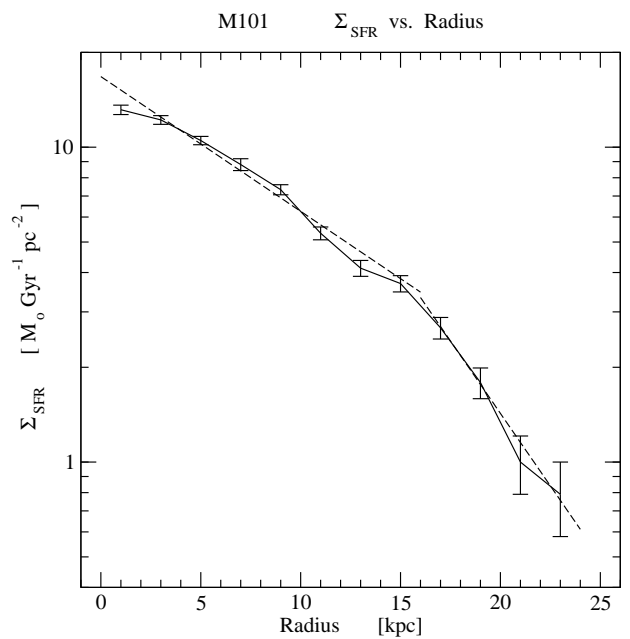}
\caption{Radial variation of the mean star formation rate per unit area,
$\SFRaj$, in 2\,kpc-wide rings in the plane of M\,101. The shape of the
curve is the same as that of the thermal emission in Fig.~\ref{fig:radial}.
Errors are standard deviations. Dashed lines show the fits giving the scale
lengths of $TH$ in Table~\ref{table:Rscale}.}
\label{fig:SFRi}
\end{figure}

As thermal radio emission is free-free emission from gas ionized by massive
stars, the present-day $\SFRa$ is proportional to the thermal surface
brightness. Therefore, we evaluated the mean value of $\SFRa$ in M\,101 by
comparing the thermal surface brightness at $\lambda$\,21\,cm,
$s_{\rm 21}$, with that of M\,33, for which $\SFRa$ is known
\citep[see][table~6]{Berkhuijsen13},
\begin{equation}
\SFRa{\rm (M\,101)} = \frac{s_{\rm 21}{\rm (M\,101)}}{s_{\rm 21}\
{\rm (M\,33)}}\,\SFRa{\rm (M\,33)}.
\label{eq:sfr}
\end{equation}
At distance $D$ we have $s_{\rm 21} = S_{\rm 21}\,4\,D^2\,/\,R^2$, where
$S_{\rm 21}$ is the thermal flux density of the area within radius $R$.
With $D = 7.4$\,Mpc, $S_{\rm 21}=TH_{\rm21}=160 \pm 13$\,mJy within $R = 30$\,kpc
(calculated from $TH_{\rm 6}$ in Table~\ref{table:sint}) for M\,101 and
$D=0.84$\,Mpc, $S_{\rm 21}=420$\,mJy within $R = 5$\,kpc and
$\SFRa=3.0 \pm 0.6\,\MGpc$ for M\,33, we find
$\SFRa$(M\,101) = $2.5\pm0.2\,\MGpc$ for the area $R < 30$\,kpc
($R \la$ 14\arcmin). We then used the $\lambda$6\,cm thermal map of M\,101 to
find the mean $\SFRaj$ in the 2\,kpc-wide rings used before
\begin{equation}
\SFRaj = \frac{TH_{{\rm 6},j}}{TH_{\rm 6}}\,\SFRa({\rm M\,101}),
\label{eq:sfr-i}
\end{equation}
where $TH_{{\rm 6},j}$ and $TH_{\rm 6}$ are the mean thermal intensity for
ring $j$ and $R<30$\,kpc, respectively. We present $\SFRaj$ as a function
of radius in Fig.~\ref{fig:SFRi}. Since $\SFRa \propto TH$, the shape of the
curve is the same as that of $TH$ in Fig.~\ref{fig:radial}. The thermal
emission from M\,101 has a systematic error of 20\% because of the
uncertainty in $\alphan$ (see Sect.~\ref{subsubsect:alphan}); therefore,
$\SFRa({\rm M\,101})$ and $\SFRaj$ also have a systematic error of 20\%.
The thermal emission and $\SFRa$ of M\,33, however, do not contain such a
systematic error because they were both derived from extinction-corrected
H$\alpha$ data \citep{Tabatabaei07b, Berkhuijsen13}.

In M\,101 the values of $\SFRaj$ range from nearly $14\,\MGpc$ at
$R<2$\,kpc to about $0.8\,\MGpc$ at $R=22-24$\,kpc, which is in good agreement
with the range derived by \citet[][fig.~1]{Zasov06} from UV and FIR data.
\citet[][fig.~8a]{Suzuki10} found values of $5-100\,\MGpc$ in spiral arms,
and the map of $\SFRa$ of \citet[fig.~20]{Leroy12} shows values of about
$16\,\MGpc$ near the centre and of $0.6\,\MGpc$ in spiral arms. Hence, the
radial distribution of $\SFRaj$ in Fig.~\ref{fig:SFRi} is consistent with
other estimates in the literature.

\begin{figure}
\centering
\includegraphics[angle=270,width=0.45\textwidth]{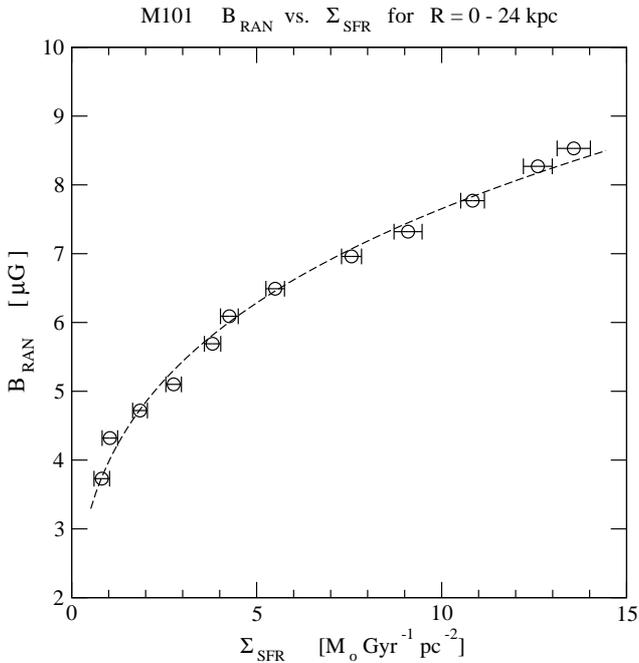}
\caption{Dependence of the turbulent magnetic field strength $\Br$ on the
star formation rate per unit area, $\SFRa$. The points represent average
values in 2\,-kpc wide rings in the plane of M\,101. The dashed line shows
the power-law fit for $R<$24\,kpc given in the text. Statistical errors
of 1 $\sigma$ are shown for $\SFRa$, but are negligble in $\Br$.}
\label{fig:BrSFR}
\end{figure}

In Fig.~\ref{fig:BrSFR} the mean values of $\Br$ in 2\,kpc\,--wide rings are
plotted against the corresponding mean values of $\SFRa$ for
$R = 0-24$\,kpc. A power-law fit to the points yields
\begin{equation}
\Br = (3.98\pm0.12)~ \SFRa^{\,\,\,0.28\pm0.02}.
\label{eq:Br}
\end{equation}
The uncertainty in $\alphan$ causes a systematic error in the exponent of
$\le 0.02$.

By using the values in 2\,kpc-wide rings, our fit refers to a correlation
on large scales. In spite of this, the exponent of $0.28\pm0.02$ is in
good agreement with those found on small scales by \citet{Chyzy08} and
\citet{Tabatabaei13b}, who derived exponents of $0.26\pm0.01$ for NGC\,4254
and $0.16\pm0.01$ for NGC\,6946, respectively, using pixel-to-pixel
correlations. The small exponent found for NGC\,6946 is attributed to
the fast cosmic ray diffusion in this galaxy.


\begin{table}
\centering
\caption{Power-law exponents $b$ in $B \propto \SFRa^b$ from
$b=L_{\rm SFR}\,/\,L_{\rm B}$. Systematic errors in $b$ due to the
uncertainty in $\alphan$ are smaller than the statistical errors}
\begin{tabular} {lcc}
\hline\hline
Field type   &  $R=0-16$\,kpc    &   $R=16-24$\,kpc       \\
\hline
$\Bt$        &  $0.22 \pm 0.03$   &   $0.24 \pm 0.05$      \\
$\Br$        &  $0.30 \pm 0.03$   &   $0.24 \pm 0.07$      \\
$\Bo$        &       -----        &   $0.22 \pm 0.03$      \\
\hline
\end{tabular}
\label{table:b}
\end{table}

As discussed above, the radial distributions of magnetic field strength in
Fig.~\ref{fig:Brad} show a break near $R=16$\,kpc causing different scale
lengths for $R<16$\,kpc and $R>16$\,kpc. We calculated the exponent $b$
in $B \propto \SFRa^b$ from the scale lengths at $R<16$\,kpc and
$R=16-24$\,kpc, given in Table~\ref{table:Rscale}, as
$b=L_{\rm SFR}/L_{\rm B}$, where $L_{\rm SFR}=L_{\rm TH}$. As can be seen
in Table~\ref{table:b}, the values of $b$ agree within errors. Although the
power law between $\Br$ and $\SFRa$ at $R<16$\,kpc may be somewhat steeper
than that at $R=16-24$\,kpc, the fit for $R=0-24$\,kpc shown in
Fig.~\ref{fig:BrSFR} with $b=0.28\pm0.02$ is within errors for both radial
ranges. At $R\,>\,$16\,kpc, $\Bo$ is also correlated with
$\SFRa$, which is not the case in NGC\,4254 \citep{Chyzy08} and NGC\,6946
\citep{Tabatabaei13b}. However, these authors used pixel-to pixel
correlations for the whole galaxy, in which a possible weak dependence
in the outer part may have been lost.

Since the total magnetic field contains a large random fraction, $\Bt$
is correlated with $\SFRa$ as well, but with a somewhat smaller exponent
than $\Br$ (see Table~\ref{table:b}). This is also the case in
NGC\,6946 \citep{Tabatabaei13b}. Furthermore, significant correlations
between the global values of $\Bt$ and $\SFRa$ have been found for a small
sample of Local Group dwarfs with $b=0.30\pm 0.04$ \citep{Chyzy11}, for
17 low-mass, Magellanic-type and peculiar galaxies with $b=0.25\pm 0.02$
\citep{Jurusik14}, for a sample of 17 galaxies with $b=0.30\pm 0.02$
\citep{Heesen14}, and for a sample of 20 nearby spiral galaxies with
$b=0.19\pm 0.03$ \citep{vanEck15}. It would be interesting to see
if the observed variation in the exponent $b$ could be related to the
considerable variation in the dependence of the local star formation rate on
the total gas surface density \citep{Bigiel08}, on variations in the
dependence of $\Bt$ on the total gas volume density, and/or on variations
in cosmic ray diffusion (fast diffusion causes a small exponent).





\subsection{Rotation measures and depolarization}
\label{subsect:RMDP}

\begin{figure}
\centering
\includegraphics[width=0.45\textwidth]{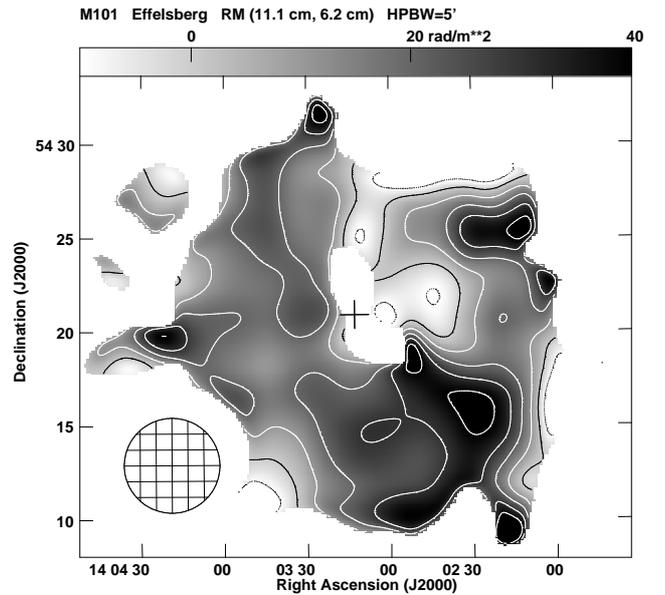}
\caption{Distribution of Faraday rotation measure $RM$(11,6) (greyscale and
contours) in M\,101 between $\lambda$\,11.1\,cm and $\lambda$\,6.2\,cm. The
data are convolved to a common beamwidth of 5\arcmin\, shown in the lower
left corner. Contour levels are -10, 0, 10, 20, 30, and $40\radm$. The
uncertainty in $RM$(11,6) is about $10-15\radm$. The cross shows the centre
of M\,101.}
\label{fig:RM}
\end{figure}

In Fig.~\ref{fig:RM} we present the distribution of the Faraday rotation
measures between $\lambda$\,11.1\,cm and $\lambda$\,6.2\,cm, $RM$(11,6).
After smoothing the $PI$(6\,cm) map to the 5\arcmin\, beamwidth of the
$PI$ map at 11\,cm, $RM$(11,6) was calculated for all data points above
2.3 times the noise in both maps. The ambiguity of $367\,\radm$ does
not influence these results. East of the major axis $RM$(11,6) varies
smoothly around $20\radm$, but in the western part strong gradients in
$RM$(11,6) occur. A comparison with Fig.~\ref{fig:therm} shows that $RM$(11,6)
is not correlated with the thermal emission from ionized gas that mainly
originates from discrete \HII regions with small volume filling factors.
Only the maximum in $RM(11,6) > 40\radm$ near the south-western major axis
coincides with intense thermal emission. Hence, $RM$(11,6) arises in the
diffuse ionized gas in M\,101. This is also the case in M\,31
\citep{Berkhuijsen03} and M\,51 \citep{Fletcher11}.

\begin{figure}
\centering
\includegraphics[width=0.45\textwidth]{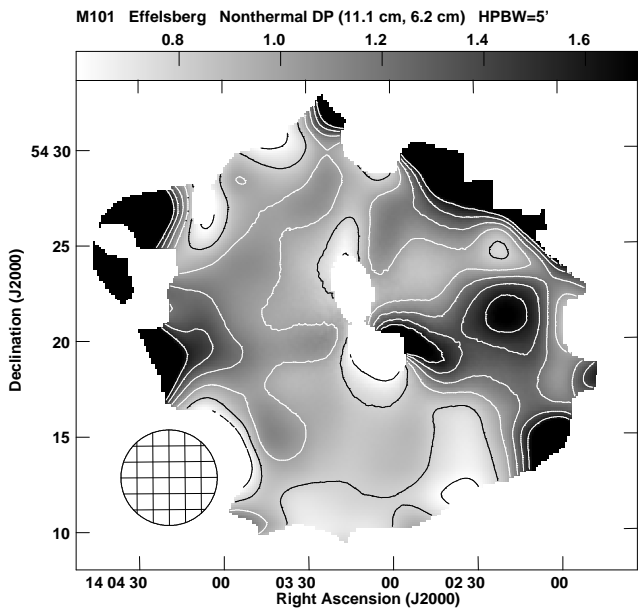}
\caption{Distribution of the non-thermal depolarization,
$\DPn$(11,6)=$\pn$(6)/$\pn$(11), in M\,101. Contour levels are 0.6, 0.8,
1.0, 1.2, 1.4, and 1.6. The uncertainty in $\DPn$(11,6) increases from 0.1
near the centre to 0.3 in the outer parts. The angular resolution is 5\arcmin.}
\label{fig:DP}
\end{figure}

The ratio of the non-thermal degree of polarization at $\lambda$\,11.1\,cm
and $\lambda$\,6.2\,cm yields the Faraday depolarization between these
wavelengths, $\DPn$(11,6) = $\pn$(11) / $\pn$(6), as the
wavelength-independent polarization cancels. The uncertainty in $\alphan$
causes a systematic error of 20\% in $\pn$(6), 12\% in $\pn$(11), and 10\%
in $\DPn$(11,6). The distribution of $\DPn$(11,6) across M\,101 is shown in
Fig.~\ref{fig:DP}. $\DPn$(11,6) generally is close to unity, varying between
about 0.7 and 1.3. This means that depolarization by Faraday effects is small.
In Sect.~\ref{subsubsect:depol} we estimate which depolarization mechanisms
are important in M\,101.

\begin{figure}
\centering
\includegraphics[width=0.45\textwidth]{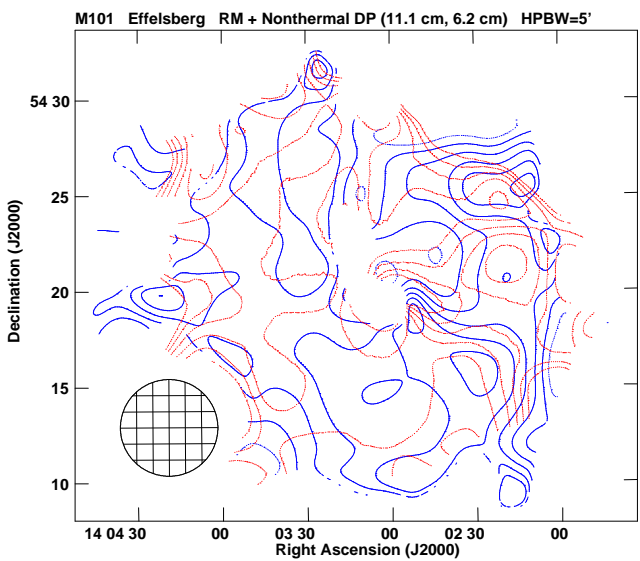}
\caption{Contours of $RM$(11,6) (thick blue lines) and $\DPn$(11,6)
(thin red lines) in M\,101 superimposed. Thin and thick lines tend to
be perpendicular to each other at their crossing points. The beamwidth
is 5\arcmin.}
\label{fig:perp}
\end{figure}

In comparing Fig.~\ref{fig:RM} and Fig.~\ref{fig:DP} one gets the impression
that contour levels of $RM$ are often perpendicular to contour levels of
$\DPn$. This is especially clear in Fig.~\ref{fig:perp} where both contour
sets are shown. Contours of $RM$ and $\DPn$ tend to be perpendicular to each
other at their crossing points. This suggests that gradients in $RM$ are a
significant cause of Faraday depolarization. This phenomenon was also observed
in M\,51 \citep{Horellou92} and M\,31 \citep{Berkhuijsen03}.

\subsubsection{Depolarization mechanisms in M\,101}
\label{subsubsect:depol}

\begin{figure}
\centering
\includegraphics[width=0.45\textwidth,angle=270]{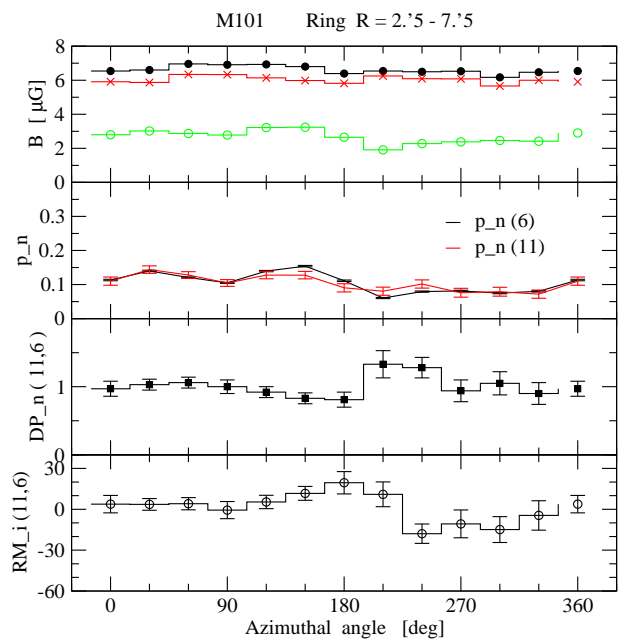}
\caption{Variation with azimuthal angle in the plane of M\,101 of the mean
value of several variables, calculated in 30\degr-wide sectors in the radial
ring $R=2\farcm5-7\farcm5$. The azimuthal angle is counted counter clockwise
from the northern major axis.
{\it Top panel}: Equipartition magnetic field strengths $\Bt$ (black dots),
$\Br$ (red crosses) and $\Bo$ (green circles). {\it Upper middle panel}:
Non-thermal polarization percentages $\pn$(6) and $\pn$(11).
{\it Lower middle panel}: Non-thermal depolarization $\DPn$(11,6).
{\it Bottom panel}: Intrinsic rotation measure $\RMi$(11,6). All error
bars are statistical errors of one $\sigma$. The uncertainty in $\alphan$
causes systematic errors of 17\% in $\Bo$, 20\% in $\pn$(6), 12\% in $\pn$(11),
and 10\% in $\DPn$(11,6).}
\vspace{0.42cm}
\label{fig:DPaz1}
\end{figure}

\begin{figure}
\vspace{0.11cm}
\centering
\includegraphics[width=0.427\textwidth,angle=270]{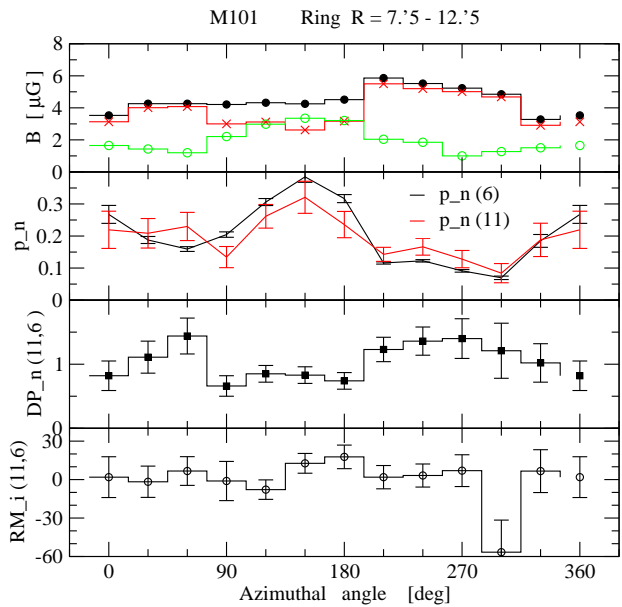}
\caption{Same as Fig.~\ref{fig:DPaz1} for the radial ring
$R=7\farcm5-12\farcm5$ in M\,101.}
\label{fig:DPaz2}
\end{figure}

In order to understand which mechanisms are causing the depolarization in
M\,101, we calculated the mean values of $\DPn$(11,6) in 30\degr-wide
sectors in two radial rings in the plane of the galaxy: an inner ring at
$R=2\farcm5-7\farcm5$ (=\,5.4\,--\,16.1\,kpc) and an outer ring at
$R=7\farcm5-12\farcm5$ (=\,16.1\,--\,26.9\,kpc). For the same sectors, we
also calculated the mean values of $\Bt, \Br$ and $\Bo$, of
$\pn$(6) and $\pn$(11), and of the intrinsic rotation measure,
$\RMi$(11,6)\,= $RM$(11,6)\,--\,$RM_{\rm f}\,$, where $RM_{\rm f}\,$ is the
rotation measure of the Galactic foreground. We estimated
$RM_{\rm f}=15\pm5\radm$ from the mean $RM$(11,6) in the two rings. The
rotation measures of the three polarized point sources located within
30\arcmin\, from the centre of M\,101 vary between $2\pm10\radm$ and
$9\pm6\radm$ \citep{Oppermann12}, which is in fair agreement with our value of
$RM_{\rm f}$. The azimuthal profiles for the two rings are shown in
Fig.~\ref{fig:DPaz1} and Fig.~\ref{fig:DPaz2}.


The profiles for the inner ring (Fig.~\ref{fig:DPaz1}) show little
variation with azimuth. The non-thermal degrees of polarization $\pn$(6)
and $\pn$(11) are nearly the same and $\DPn$(11,6) remains close to 1.
Hence, Faraday depolarization is unimportant and the low values of
$\pn\simeq 0.1$ must be due to wavelength-independent polarization
\footnote{Instead of wavelength-independent depolarization we use
the more accurate description of wavelength-independent {\it polarization}
(see \citet{Sokoloff98}), emerging from ordered fields (at small wavelengths)
or from sheared or compressed random magnetic fields in the emission region
(see Sect.~\ref{subsubsect:holes}).}.
The top panel shows that in all sectors $\Br$ dominates as $\Br/\Bo\simeq 2.5$
and $\Br/\Bt\simeq 0.9$. $\RMi$(11,6) (bottom panel) is generally small,
but changes from $\simeq 15\radm$ to $\simeq -15\radm$ between
$Az=210\degr$\, and $Az=240\degr$. Figure~\ref{fig:RM} shows a
strong gradient in $RM$(11,6) in these sectors, which  causes the depression
in $PI$ south-west of the centre in Fig.~\ref{fig:pipc6}. This area is
coincident with an extended minimum in the \HI\,\,map of \citet{Braun95}.

In the outer ring (Fig.~\ref{fig:DPaz2}) the situation is more complex.
From $Az=90\degr$ to $Az=180\degr$, the non-thermal polarization
percentages are increased and show a pronounced maximum at $Az=150\degr$.
In these sectors $\pn{\rm (11)} < \pn{\rm (6)}$ and $\DPn{\rm(11,6)} < 1$,
indicating Faraday depolarization. In the same interval the ordered field
strength $\Bo$ is increased and $\Br/\Bo$ has dropped to $\simeq 1$. In sector
$Az=210\degr$ $\Br$ suddenly increases by 2\muG. This is caused by the large
\HII\,\,complex south-west of the nucleus that is visible as a bright source
in both thermal and non-thermal  intensity (i.e. see Fig.~\ref{fig:therm}).
$\RMi$(11,6) is small in all sectors ($<|20|\radm$), apart from the sector
at $Az=300\degr$ where it is strongly negative with nearly $-60\radm$. This
sector contains a strong decrease in $RM$(11,6) around
RA=$14^h\,02^m\,45^s$, DEC=54\degr\,28\arcmin\,35\arcsec (see Fig.~\ref{fig:RM}).
Here $\pn$(6) and $\pn$(11) reach a minimum of less than
0.1 and $\Br/\Bo$ becomes $\simeq 4$. The minimum in the polarization
degrees is due to wavelength independent polarization as
$\pn$(6)\,$\simeq\,\pn$(11).


We discuss the wavelength-independent polarization in the inner ring in the
next section. Here we estimate whether Faraday depolarization could explain
$\pn$(6) and $\pn$(11) in the sector at $Az\,=\,150$\degr in the outer ring,
where $\DPn{\rm (11,6)}\simeq 0.8$ (see Fig.~\ref{fig:DPaz2}).

Internal Faraday dispersion usually is the strongest Faraday
effect, for which
\citet{Burn66} and \citet{Sokoloff98} give the expression
\begin{equation}
\pn(\lambda) = p_{\rm 0}\,(1-\exp{(-2S)})\,/\,2S,
\label{eq:fardisp}
\end{equation}
where $S=\sigma_{\rm RM}^2\,\lambda^4$ and $p_{\rm 0}=0.75$ is the maximum
degree of polarization
\footnote{$p_{\rm 0} = (1+\alphan) / (5/3 + \alphan) = 0.74 \pm 0.09$ for
$\alphan = 0.92$.};
$\sigma_{\rm RM}$ is the standard deviation of the
intrinsic rotation measure $\RMi$. For the wavelengths of
$\lambda\lambda$\,6.2\,cm and 11.1\,cm, we find that $\sigma_{\rm RM}=40\radm$
gives the observed value of $\DPn$(11,6)\,$\simeq0.8$. This value of
$\sigma_{\rm RM}$ is similar to those in NGC\,6946 \citep{Beck07} and IC\,342
\citep{Beck15} of $38\radm$ and $55\radm$, respectively.

Although $\sigma_{\rm RM}=40\radm$ can explain $\DPn=0.8$, the values of
$\pn$(6)\,=\,0.73 and $\pn$(11)$\,\,=\,0.50$ resulting from
Eq.~\ref{eq:fardisp}, are much higher than those observed, which are
$\pn$(6)$\,\,=\,0.39$ and $\pn$(11)$\,\,=\,0.31$. Therefore, the value of
$\pn\simeq 0.40$ is the result of wavelength-independent polarization. This
rather high value could partly come from anisotropic magnetic fields
\citep{Fletcher11} (see Sect.~\ref{subsubsect:holes}).
Thus in the sector $Az=150$\degr\, in the outer ring the combination of
Faraday dispersion and wavelength-independent polarization can explain the
observations, where the latter is the dominant polarization mechanism.

It is interesting to see whether the value of $\sigma_{\rm RM}=40\,\radm$
is consistent with the properties of the magneto-ionic medium in M\,101.
We can estimate $<\ne>$, the average electron density along the line
of sight (in $\ccm$), using the relation
\begin{equation}
\sigma_{\rm RM}\,=\,0.81\,<\ne>\,\Bra\,\sqrt{L_{\rm ion}\,d\,/\,f},
\end{equation}
where $\Bra$ is the strength of the component of the isotropic random
field along the line of sight (in \muG); $L_{\rm ion}$ is the path length
through the layer of diffuse gas (in pc) containing ionized cells with a
typical size of $d$  = 50\,pc, which is the coherence length of turbulence in
the ISM \citep{Ohno93, Berkhuijsen06}; and $f$ is their volume filling
factor along $L_{\rm ion}$. For an exponential scale height of the ionized
layer of 1\,kpc, $L_{\rm ion} = 2000\,/\,\cos(i)$ = 2300\,pc, where we
assume that we see polarized emission from both sides of the disk. With
$\Bra=\Br\,\sqrt{1/3}\,=\,1.6$\,\muG\, and $f = 0.5$ \citep{Berkhuijsen06}
we find $<\ne>\,=\,0.06\,\ccm$, which is about three times higher than found
near the sun \citep{Berkhuijsen08}. However, a smaller filling factor or a
larger size of the ionized cells would bring $<\ne>$\, closer to the MW value.

Alternatively, we may estimate $<\ne>$ from the maximum intrinsic rotation
measure in the outer ring using
\begin{equation}
\RMi\,=\,0.81\,<\ne>\,\Boa\,L_{\rm ion},
\end{equation}
where $\Boa\,=\,\Bo\,\sin(i)$ is the strength of the ordered magnetic field
component along the line of sight, assumed to be regular.
With $|\RMi| = 18\,\radm$, $\Bo = 3.1$\,\muG\, (see Fig.~\ref{fig:DPaz2}),
and $L_{\rm ion}$ = 2300\,pc, we have $<\ne>\, = 0.006\,\ccm$, which is about
one-third of the value near the sun. The difference between the two estimates
of $<\ne>$\,suggests that the observed polarized emission mainly travels
through thin, diffuse ionized gas, whereas the depolarization by Faraday
dispersion is mainly caused by the denser, ionized clouds. However, we should
regard this low value of $<\ne>$\, as a lower limit if part of the ordered
field observed in polarized emission is anisotropic (sheared or compressed
field), which does not contribute to Faraday rotation and $\RMi$
\citep{Fletcher11}.

We conclude that the low degrees of polarization in M\,101 are mainly caused
by dispersion of polarization angles by random magnetic fields in the
emission regions, leading to wavelength-independent polarization. Faraday
dispersion also plays a role, but only in some regions.


\subsubsection{{\bf What causes wavelength-independent polarization?}}
\label{subsubsect:holes}

In the foregoing Section, we showed that wavelength-independent
polarization is the main  polarization mechanism in M\,101. We now
estimate under which circumstances wavelength-independent polarization
with $\pn= 0.1$ is obtained in the inner ring
($R= 2\farcm5$ - $7\farcm5$) (Fig.~\ref{fig:DPaz1}).

In M\,51, \citet{Fletcher11} found that most of the polarized radio
emission from the disk, observed at short wavelengths, arises from
anisotropic random magnetic fields that do not contribute to the rotation
measure. This may be a general property of galaxy disks, and the low values
of $\RMi$ in M\,101 suggest that anisotropic magnetic fields may be
important. Therefore, we discuss two possibilities (recall footnote 2 to
Sect.~\ref{subsubsect:radialB}):\\
(a) polarization by an ordered (regular and/or anisotropic random) field; and\\
(b) polarization by a purely anisotropic random field.

For case (a), we use Eq.\,(24) of \citet{Sokoloff98} for
wavelength-independent polarization in a partly ordered magnetic field,
\begin{equation}
\pn = p_{\rm 0}~\frac{\Bop^2}{\Bop^2 +\sigma_{\rm r}^2},
\label{eq:lamind}
\end{equation}
where $\pn$ is the observed non-thermal degree of polarization,
$p_{\rm 0}=0.75$ the maximum intrinsic degree of polarization,
$\Bop=\Bo\,\cos(i)$ the strength of the ordered magnetic field component in
the sky plane, and $\sigma_{\rm r}$ the {\it standard deviation} of the
isotropic random field in the plane of the sky. As $\sigma_{\rm r}$ originates
from a large number of turbulent cells with a typical size of 50\,pc, it does not
contribute to the polarized emission. With $\pn$ and $\Bo$ known, we can
calculate $\sigma_{\rm r}$.
\footnote{Equation~(\ref{eq:lamind}) is valid for a constant value of the
density of cosmic ray particles. In case of energy equipartition between
cosmic ray particles and magnetic fields, $\pn$ is about $25\,\%$ higher
than given by Equation~\ref{eq:lamind} \citep{Sokoloff98} and
$\sigma_{\rm r}$ is about $15\,\%$ higher than the value we derived
here.}

In the inner ring, $\Bo\simeq2.8$\muG (see Fig.~\ref{fig:DPaz1}), and the
observed value of $\pn\simeq0.1$ is reached for $\sigma_{\rm r}\,=\,5.7$\muG.
As the strength of the random field in the plane of the sky is
$\Brp=\Br \sqrt{2/3}\,=\,4.9$\muG, the standard deviation in $\Br$ is about
the same as $\Br$ itself, $\sigma_{\rm r}\,\simeq\,1.2\,\Br$. If part of
$\pn$ is due to anisotropic random fields, as addressed in case (b),
$\sigma_{\rm r}$ is larger than obtained here.

In case (b), we assume that the small degree of polarization in the
inner ring entirely emerges from anisotropic random fields \citep{Sokoloff98},
consisting of many elongated cells causing polarized emission but no rotation
measure. This means that the magnetic fields in the emission regions are
tangled and disrupted by, for example, stellar winds, supernova shocks,
expanding shells, gas outflow from star formation complexes, and Parker loops.
By estimating the typical size of the `cells' of field irregularities, we may
identify the main cause of the wavelength-independent polarization.

For a random distribution of polarization angles the number of cells N
can be found from \citep{Beck99}
\begin{equation}
\pn = p_{\rm 0}\,N^{-0.5},
\end{equation}
where $\pn$ is the observed non-thermal degree of polarization, $p_0$ = 0.75
the maximum degree of polarization in an undisturbed regular magnetic field,
and N the number of cells in the volume observed. In the inner ring,
$\pn\,\simeq\,0.1$ in most sectors (see Fig.~\ref{fig:DPaz1}), giving
$N\simeq60$ in the sector volume. This is a lower limit for $N$ since we know
that part of the polarized emission must come from regular magnetic fields
causing the observed $\RMi$.

If a 30\degr-\,wide sector contains $N$ cells of size $d$ and a volume
filling factor $f_{\rm v}$, we obtain
\begin{equation}
N = \frac{f_{\rm v}\,L_{\rm em}\,((R_2)^2 - (R_1)^2)\,/\,12} {4/3\,(d/2)^3},
\end{equation}
yielding
\begin{equation}
d = [f_{\rm v}\,L_{\rm em}\,((R_2)^2 - (R_1)^2) / 2N]^{1/3},
\end{equation}
where $L_{\rm em}$ is the line of sight through the emission region and
$R_2$ and $R_1$ are the radii determining the inner ring.

The volume filling factor is a combination of the area filling factor
$f_{\rm a}$ and the filling factor along the line of sight
$f_{\rm L} = d\,/\,L_{\rm em}$, $f_{\rm v} = f_{\rm L} f_{\rm a}$. Inserting
this into the above equation and solving for $d$ again yields
\begin{equation}
d = f_{\rm a}^{0.5}\,[((R_2)^2 - (R_1)^2)\,/\,2\,N]^{0.5}.
\label{eq:size0}
\end{equation}
Thus $d$ is independent of $L_{\rm em}$ and directly proportional to
$N^{-0.5}$. With $R_2$\,=\,16.1\,kpc, $R_1$\,=\,\,5.4\,kpc, and $N\,=\,60$,
we find
\begin{equation}
d \simeq  f_{\rm a}^{0.5}\,1400\,{\rm pc}.
\label{eq:size1}
\end{equation}
As $f_{\rm a} < 1$, the typical size of the cells of field irregularities
responsible for the wavelength-independent polarization may be of the
order of 1\,kpc, which is much larger than the typical size of 50\,pc of
supernova remnants. Instead they could be large shells caused by multiple
supernova explosions, chimneys of gas rising from star-forming regions, or
Parker loops. The frequency and size of chimneys and Parker loops in galactic
disks are poorly known \citep{Mao15}, but large shells and superbubbles have
been observed in many galaxies \citep{Bagetakos11} and could be an important
cause of disordered magnetic fields in galactic disks.

In M\,101 \citet{Kamphuis93} detected 52 \HI\,\,shells, visible as holes
in the \HI\, column density distribution and in position--velocity diagrams.
Their diameters range from about 700\,pc to about 2500\,pc, but many shells
below and around the resolution limit of about 500\,pc have been
missed. In a similar study on NGC\,6946 with slightly better resolution,
\citet{Boomsma08} estimated that at least two-thirds of the shells with sizes
above the resolution limit had not been detected. If this also holds for
M\,101, it should at least contain 156 shells with diameters between
500\,pc and 2500\,pc. This still is a lower limit because well-resolved
studies on M\,31 \citep{Brinks86} and M\,33 \citep{Deul90} show that
the size distribution of \HI\,\,shells peaks at $200-300$\,pc (see also
\citet{Bagetakos11}).

\citet{Kamphuis93} calculated the area filling fraction $f_{\rm a}$ of
the observed shells as a function of radius. In the inner ring at
$R$ = 5.4\,--\,16.1 \,kpc, the mean value of $f_{\rm a}$ = 0.16, which
may increase to $f_{\rm a}$ = 0.2 if the missing smaller shells are added.
Inserting this into Eq.~\ref{eq:size1}, we find a mean size of the shells
causing the wavelength-independent polarization of $d$ = 625\,pc.

However, $d$ = 625\,pc is an upper limit because part of the observed
polarized emission, and thus of $p_{\rm n}$, must come from regular
magnetic fields observed as Faraday rotation measures (Fig.~\ref{fig:DPaz1}).
If, for example, only half of the polarized emission were due to
anisotropic random magnetic fields, reducing $p_{\rm n}$ to 0.05, the number
of cells in a sector would increase to $N = 225$ and their mean diameter
would decrease to $d = 320$\,pc. This diameter comes close to the most common
size of \HI\,shells in well-resolved galaxies.

We conclude that the wavelength-independent polarization in M\,101
could partly be due to strong disturbances of the regular magnetic
field by explosive events that give rise to \HI\,shells with mean diameters
of less than 625\,pc.


\subsection{Large-scale magnetic field and \HI\,\,spiral arms in M\,101}
\label{subsect:pitch}

In this section we investigate the relationship between the large-scale,
ordered magnetic field in M\,101 and spiral arms seen in H$\alpha$ and
\HI\,by studying the orientation of the magnetic pitch angles, which we calculate
from the polarization angles at $\lambda$\,6.2\,cm.

Figure~\ref{fig:RM} shows that Faraday rotation in M\,101 generally does not
exceed $30\radm$, yielding a maximum rotation angle of 7\degr\ at $\lambda$\,
6.2\,cm. This rotation angle is comparable to the uncertainty in the
polarization position angle, and it only exceeds $40\radm$ (corresponding to
9\degr) in small isolated regions. Therefore, we did not correct the
polarization angles for the Faraday rotation offset.

\begin{figure}
\centering
\includegraphics[width=0.45\textwidth]{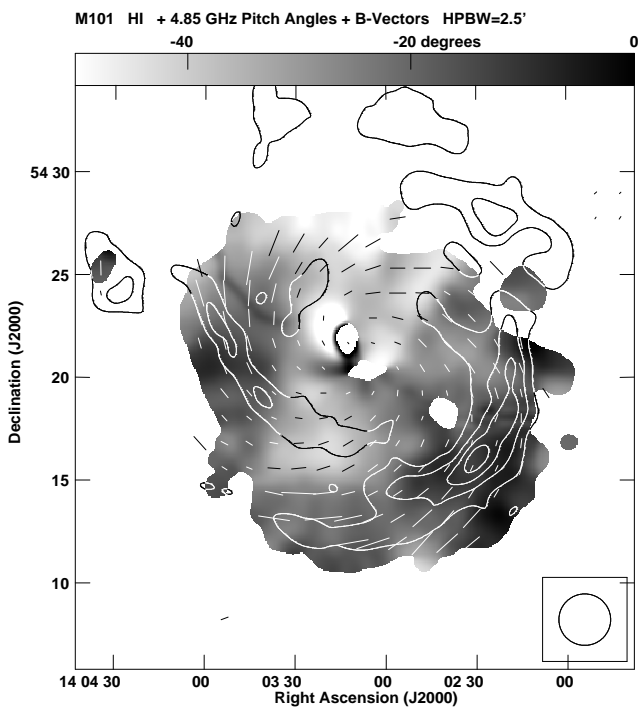}
\caption{The magnetic pitch angles in M\,101 at $\lambda$\,6.2\,cm shown  in
greyscale with apparent B-vectors and some contours of the \HI\, spiral
arms \citep{Braun95} superimposed, smoothed to the beamwidth of $2\farcm5$.
The contour levels are (10, 15, 20, and 25)\,$10^{20}\,{\rm cm^{-2}}$. The
pitch angles are corrected for the inclination of
30\degr, but not for Faraday rotation because it is $<7\degr$.
The pitch angles are smallest in the south-west, where the apparent B-vectors
run nearly parallel to the \HI\,arms, but  pitch angles near $-40\degr$ occur
in the south and the north. The length of the B-vectors is proportional to the
non-thermal degree of polarization with $1\arcmin\,=20\%$.}
\label{halpit}
\end{figure}

Figure~\ref{halpit} shows the large-scale distribution of magnetic pitch angles
(defined as the angle between the apparent B-vectors at $\lambda$\,6.2\,cm,
corrected for the inclination, and the local azimuthal direction in the disk)
and the brightest parts of the \HI\,spiral arms containing star-forming regions.
Generally, the pitch angles of the B-vectors are largest in the northern and
smallest in the SW region of M\,101. They are also very small in the outer part
of the southern disk, but become larger towards the centre. While the SW region
shows a radial {\it decrease} of the magnetic pitch angle, the magnetic pitch
angle tends to {\it increase} with radius towards the NW, N, and NE.

There is no clear association of magnetic pitch angles with the star
formation distribution in M\,101 (compare also Fig.~\ref{fig:therm}).
Large pitch angles are found at the position of the extended star-forming
complex south of the disk centre as well as in a quiescent outer northern
region. There is no clear indication that the ratio of radial-to-azimuthal
magnetic field is enhanced close to star-forming regions. One might expect
this if the radial field is produced locally out of the azimuthal field by a
turbulent dynamo boosted by turbulent activity in actively star-forming
portions of the disk.

\begin{figure*}
\vspace{1.2cm}
\hspace{1.2cm}
\includegraphics[width=13cm]{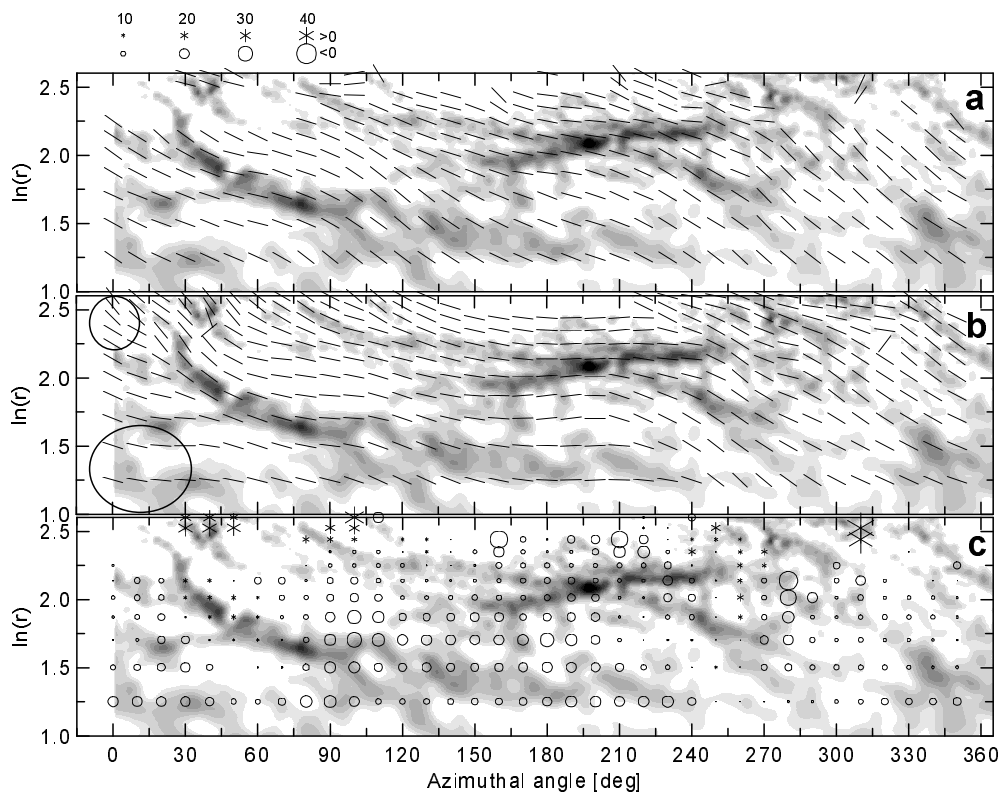}
\caption{{\bf a)}\, Orientations of the apparent B-vectors (plotted with
equal length) in M\,101 at $\lambda$\,6.2\,cm (corrected for inclination) in
the frame of azimuthal angle in the disk plane and $ln(r)$ (r is the galacto-
centric radius in arcmin), overlaid onto the \HI\,distribution of
\citet{Braun95}, smoothed to a beamwidth of 20\arcsec.
{\bf b)}\, Orientation of the B-vectors (plotted with equal length) of
the magnetic field model assuming the magnetic field to be parallel to the \HI\,
filaments. Ellipses show the beamwidth at selected radii.
{\bf c)}\, Differences between model and observations shown as symbols with
size proportional to the pitch angle difference (the scale is given at the
top of the Figure). As the pitch angles are mostly negative, a negative
difference means a more inclined B-vector in the observed than in the model
map. The azimuthal angle runs counter clockwise from the NE major axis.}
\label{pit-all}
\end{figure*}

We checked the hypothesis that the magnetic field orientations in the disk
of M\,101 may be controlled by compression effects in \HI\,filaments by
comparing the orientations of the B-vectors at $\lambda$\,6.2\,cm to those
of \HI\,structures \citep{Braun95}. Figure~\ref{pit-all}a shows that in the
frame of azimuthal angle in the disk -- $ln(r)$, where r is the galacto-centric
radius, the magnetic field orientations generally follow the \HI\,filaments.
The long, weak filament starting at $Az=70\degr$, $ln(r)=2.5$ runs parallel to
B-vectors. At $Az=150\degr-210\degr$, $ln(r)>2,$ the filament and magnetic field
become nearly azimuthal and then they both bend towards smaller $ln(r)$. On the
other hand, the B-vectors are apparently inclined with respect to a bright
filament crossing the former filament at $Az=210\degr$, $ln(r) = 2.1$. The
magnetic field orientations are well aligned with a prominent \HI\,filament at
$Az=30\degr-90\degr$, beyond which they become inclined to the more diffuse
continuation of this filament. At $Az>270\degr$, $ln(r)<2.1,$ there is good
agreement between large magnetic pitch angles and the general preponderance of
highly inclined, diffuse \HI\, structures.

Because of significant beam-smearing effects, the question of an agreement or
disagreement of observed magnetic pitch angles and the orientations of
\HI\,structures must be solved by means of beam-smoothed models of polarized
emission. For this purpose, we digitized the loci of maxima of clearly
identifiable \HI\,filaments. We ignored localized wiggles that are much smaller
than our beam, and traced a general trend of each filament. Each filament was
split into segments about 1\arcmin\, long. The magnetic field was assumed
to run locally parallel to each segment.
We integrated contributions from particular segments to the Q and U Stokes
parameters convolved to a beam of $2\farcm4$ using techniques described
by \citet{Urbanik97} and \citet{Soida96}. To best reproduce the observations,
the contributions from particular filament segments were weighted with the
polarized intensity observed at this position. Then, we combined the Q and U
distributions to a map of polarization angles, and analysed these angles in
the same way as the observed angles. As a result of the smoothing procedure
polarization angles between \HI\, filaments also occur.

\begin{figure}
\centering
\includegraphics[width=0.45\textwidth]{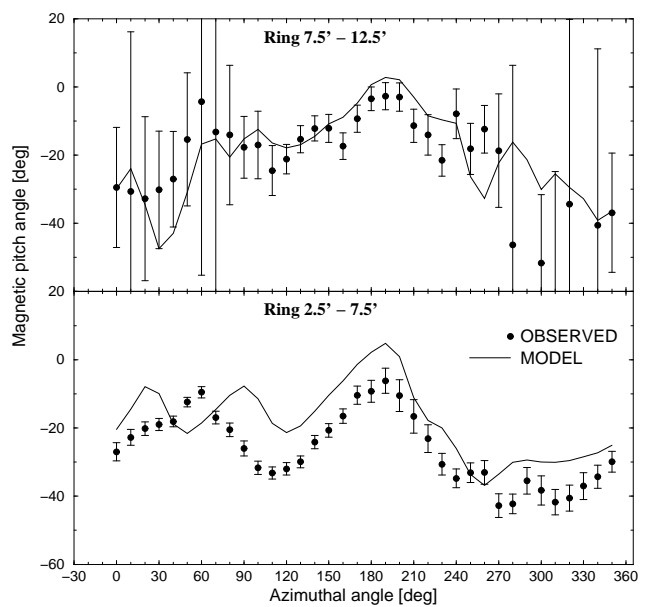}
\caption{Variations with azimuthal angle of the magnetic pitch angle in
M\,101 averaged in 10\degr\, sectors in the rings $R=2\farcm5-7\farcm5$ and
$R=7\farcm5-12\farcm5$, compared to the variations expected if the
magnetic field is aligned with \HI\, of the model filaments. The azimuthal
angle is counted counter-clockwise from the NE major axis. The pitch angles
are corrected for the inclination as well as for the Faraday rotation.}
\label{wide}
\end{figure}

The results are compared to observations in Fig.~\ref{pit-all}\,a, b,
and c. The \HI\,filaments and apparent B-vectors show similar large-scale
variatons in the pitch angles.
The model well reproduces the magnetic field following the long filament
running from $Az=70\degr$, $ln(r)=2.5$ to $Az=270\degr$\, close to the
centre, and another filament at $Az=30\degr-90\degr$. The preponderance of highly
inclined B-vectors at azimuths $Az>270\degr$, $ln(r)<2.1$ is also well
reproduced. However, large differences reaching 30\degr\,occur between
$Az=90\degr$\, and $Az=210\degr$\, at $ln(r)<2$. Here M\,101 has a
significantly inclined magnetic field, but the model predicts an almost
azimuthal field. This results from assuming that the  magnetic field is
parallel to a prominent filament at $Az=130\degr-240\degr$\, and to an
almost horizontally running fuzzy structure at $ln(r)= 1.3$. Some large
discrepancies at $ln(r)\,\ge\,2.2$ are uncertain because the signal-to-noise
ratio in $PI$ is low. Thus, in M\,101 there are some regions that show a
stronger radial magnetic field than one would expect from field lines
consistently parallel to the filaments.

The pitch angles are shown in Fig.~\ref{wide}, where corrections for
the mean Faraday rotation at given azimuthal angles have been applied. The
model and observed magnetic pitch angles vary in a similar way with
azimuthal angle, which means that the orientation of \HI\,filaments is
efficiently controlling the pitch angle of magnetic fields. There is a clear
asymmetry in the pitch angles between the northern ($Az=0\degr\,$) and
southern ($Az=180\degr\,$) major axis: The pitch angles in the north are
much more negative than in the south. This may be related to the
lopsidedness of M\,101, the disk of which is much more extended along the
northern major axis than along the southern major axis, which is possibly the
result of past encounters with members of the M\,101 group
\citep[e.g.][]{Karachentsev14, Mihos13, Jog09,Waller97}. The \HI\,filaments
may be associated with density perturbations caused by the encouter(s),
aligning the magnetic field like density waves do in spiral arms.


The model magnetic field, based on the \HI\,filaments, generally has smaller
pitch angles than are observed (Fig.~\ref{wide}). In the inner ring the
unweighted mean difference is $8\pm1\degr$. In the outer ring the difference
seems to be smaller but cannot be determined with sufficient accuracy. In the
barred spiral galaxy M\,83, the magnetic pitch angles are on average about
20\degr\ larger than the pitch angles of the material arms traced in CO and
\HI\, \citep{Frick16}.

We suggest that there is a source of radial magnetic fields other than
pure compression aligning the field lines with local \HI\,filaments.
The large-scale $\alpha-\Omega$ dynamo naturally produces the radial
magnetic field component and hence increases the magnetic pitch angle
\citep{Beck96}. The excess of the radial component is strongest in the
inter-arm space away from star-forming regions (Fig.~\ref{pit-all}). The
large-scale dynamo process apparently works more efficiently in inter-arm
regions, which could be the result of enhanced outflows in the spiral arm
regions \citep{Chamandy15}.

\section{Summary}
\label{sect:sum}

We present observations of total and polarized emission from the spiral
galaxy M\,101 at $\lambda\lambda$\,6.2\,cm and 11.1\,cm, which we carried out
with the Effelsberg telescope. The angular resolutions are
$2\farcm5$ (=\,5.4\,kpc) and $4\farcm4$ (=\,9.5\,kpc), respectively. We used
these data to study the various emission components and the properties of the
magnetic field in M\,101 . Our main results are summarized below.

-- The thermal radio emission is closely correlated with the spiral arms,
but the non-thermal emission is more smoothly distributed indicating
diffusion of cosmic ray electrons from their places of origin. The
thermal fraction at $\lambda$\,6.2\,cm is $f_{\rm th} < 45\pm6\%$ with
a systematic error of 10\%.

-- The radial distributions of thermal and non-thermal emission show a
break near $R=16$\,kpc (=$7\farcm4$), where they steepen to an exponential
scale length of $L \simeq 5$\,kpc, which is about 2.5 times smaller than at
$R<16$\,kpc. The distribution of the polarized emission has a broad maximum
near $R=12$\,kpc and beyond $R=16$\,kpc, this distribution also decreases with
$L \simeq 5$\,kpc. As near the radius of $R=16$\,kpc  the radial distribution
of the optical emission also steepens \citep{Mihos13} and the position
angle, inclination, and the velocity structure of the \HI\,gas change
\citep{Kamphuis93}, a major change in the structure of M\,101 must occur near
this radius.

-- The change in the structure of M\,101 near $R=16$\,kpc is also apparent
in the radial distributions of the magnetic field strengths $\Bt$, $\Br$,
and $\Bo$. Beyond $R=16$\,kpc their radial scale length is about 20\,kpc,
which implies that they extend to $R=70$\,kpc before decreasing to
0.3\muG. The strength of $\Bt$ ranges from nearly 10\muG at $R<2$\,kpc to
4\muG at $R=22-24$\,kpc. As the random magnetic field dominates
in M\,101 ($\Br/\Bo\simeq2.4$), $\Bo$ is weak, varying between 3\muG at
$R=12-14$\,kpc and 2.1\muG at $R=22-24$\,kpc. The mean field strengths for
$R<24$\,kpc are $\Bt=6.4$\muG, $\Br=5.9$\muG, and $\Bo=2.5$\muG.

-- The integrated thermal luminosity for $R<30$\,kpc yields a mean SFR
per unit area of $\SFRa=2.5\pm0.2\,\MGpc$, which is somewhat smaller than that
of M\,33. Averages in 2\,kpc-wide rings in the plane of the galaxy decrease
from $14\,\MGpc$ at $R<2$\,kpc to $0.8\,\MGpc$ at $R=22-24$\,kpc. $\SFRa$
and the ring averages have a systematic error of 20\%.

-- At radii $R<24$\,kpc, the random magnetic field depends on $\SFRa$
with a power-law exponent of $b=0.28\pm0.02$. The systematic error
in $b$ is $\le 0.02$.

-- In most regions in M\,101 rotation measures $RM$(11,6) are between
$-30\radm$ and $30\radm$, and the non-thermal depolarization $\DPn$(11,6)
varies between 0.7 and 1.3.

-- Wavelength-independent polarization, caused by the random magnetic
field in the emission regions, is the main polarization mechanism in
M\,101. In some areas beyond $R=16$\,kpc, Faraday dispersion also plays a role.
We show that energetic events causing \HI\,shells of several hundred pc in
diameter could be responsible for part of the wavelength-independent polarization.

-- The ordered magnetic field is generally aligned with the spiral arms
showing the same large-scale azimuthal asymmetries, caused by the
interaction of M\,101 with some of its group members.
However, a beam-smoothed model shows that the magnetic pitch angle
variations over the  disk cannot  be entirely caused by alignment of
magnetic field lines along \HI\,filaments as there are substantial local
deviations. The magnetic pitch angles are on average about 8\degr\, larger
than the pitch angles of the model \HI\,filaments, indicating the action
of a large-scale dynamo.


\begin{acknowledgements}
This research has been supported by a scientific grant from the National
Science Centre (NCN), dec. No. 2011/03/B/ST9/01859. We thank Dr. Aritra
Basu for careful reading of the manuscript and useful comments. We
also thank the anonymous referee, whose detailed comments led to several
improvements of the text.
\end{acknowledgements}


\bibliographystyle{aa}
\bibliography{m101.v180116}

\begin{thebibliography}{66}
\expandafter\ifx\csname natexlab\endcsname\relax\def\natexlab#1{#1}\fi

\bibitem[{{Bagetakos} {et~al.}(2011){Bagetakos}, {Brinks}, {Walter}, {de \
  Blok}, {Usero}, {Leroy}, \& {Rich}}]{Bagetakos11}
{Bagetakos}, I., {Brinks}, E., {Walter}, F., {et~al.} 2011, \aj, 141, 23

\bibitem[{{Beck}(2007)}]{Beck07}
{Beck}, R. 2007, \aap, 470, 539

\bibitem[{{Beck}(2015)}]{Beck15}
{Beck}, R. 2015, \aap, 578, A93

\bibitem[{{Beck} {et~al.}(1999){Beck}, {Berkhuijsen}, \& {Uyaniker}}]{Beck99}
{Beck}, R., {Berkhuijsen}, E.~M., \& {Uyaniker}, B. 1999, in Plasma Turbulence
  and Energetic Particles in Astrophysics, ed. M.~{Ostrowski} \&
  R.~{Schlickeiser}, 5--17

\bibitem[{{Beck} {et~al.}(1996){Beck}, {Brandenburg}, {Moss}, {Shukurov}, \&
  {Sokoloff}}]{Beck96}
{Beck}, R., {Brandenburg}, A., {Moss}, D., {Shukurov}, A.~., \& {Sokoloff}, D.
  1996, \araa, 34, 155

\bibitem[{{Beck} \& {Krause}(2005)}]{Beck05}
{Beck}, R. \& {Krause}, M. 2005, Astronomische Nachrichten, 326, 414

\bibitem[{{Berkhuijsen} {et~al.}(2003){Berkhuijsen}, {Beck}, \&
  {Hoernes}}]{Berkhuijsen03}
{Berkhuijsen}, E.~M., {Beck}, R., \& {Hoernes}, P. 2003, \aap, 398, 937

\bibitem[{{Berkhuijsen} {et~al.}(2013){Berkhuijsen}, {Beck}, \&
  {Tabatabaei}}]{Berkhuijsen13}
{Berkhuijsen}, E.~M., {Beck}, R., \& {Tabatabaei}, F.~S. 2013, \mnras, 435,
  1598

\bibitem[{{Berkhuijsen} {et~al.}(2006){Berkhuijsen}, {Mitra}, \&
  {Mueller}}]{Berkhuijsen06}
{Berkhuijsen}, E.~M., {Mitra}, D., \& {Mueller}, P. 2006, Astronomische
  Nachrichten, 327, 82

\bibitem[{{Berkhuijsen} \& {M{\"u}ller}(2008)}]{Berkhuijsen08}
{Berkhuijsen}, E.~M. \& {M{\"u}ller}, P. 2008, \aap, 490, 179

\bibitem[{{Bigiel} {et~al.}(2008){Bigiel}, {Leroy}, {Walter}, {Brinks}, {de
  Blok}, {Madore}, \& {Thornley}}]{Bigiel08}
{Bigiel}, F., {Leroy}, A., {Walter}, F., {et~al.} 2008, \aj, 136, 2846

\bibitem[{{Boomsma} {et~al.}(2008){Boomsma}, {Oosterloo}, {Fraternali}, {van
  der Hulst}, \& {Sancisi}}]{Boomsma08}
{Boomsma}, R., {Oosterloo}, T.~A., {Fraternali}, F.~., {van der Hulst}, J.~M.,
  \& {Sancisi}, R. 2008, \aap, 490, 555

\bibitem[{{Braun}(1995)}]{Braun95}
{Braun}, R. 1995, \aaps, 114, 409

\bibitem[{{Brinks} \& {Bajaja}(1986)}]{Brinks86}
{Brinks}, E. \& {Bajaja}, E. 1986, \aap, 169, 14

\bibitem[{{Burn}(1966)}]{Burn66}
{Burn}, B.~J. 1966, \mnras, 133, 67

\bibitem[{{Chamandy} {et~al.}(2015){Chamandy}, {Shukurov}, \&
  {Subramanian}}]{Chamandy15}
{Chamandy}, L., {Shukurov}, A., \& {Subramanian}, K. 2015, \mnras, 446, L6

\bibitem[{{Chyzy}(2008)}]{Chyzy08}
{Chyzy}, K.~T. 2008, \aap, 482, 755

\bibitem[{{Chy{\.z}y} {et~al.}(2011){Chy{\.z}y}, {We{\.z}gowiec}, {Beck}, \&
  {Bomans}}]{Chyzy11}
{Chy{\.z}y}, K.~T., {We{\.z}gowiec}, M., {Beck}, R., \& {Bomans}, D.~J. 2011,
  \aap, 529, A94

\bibitem[{{de Vaucouleurs} {et~al.}(1976){de Vaucouleurs}, {de Vaucouleurs}, \&
  {Corwin}}]{Vaucouleurs76}
{de Vaucouleurs}, G., {de Vaucouleurs}, A., \& {Corwin}, J.~R. 1976, in Second
  reference catalogue of bright galaxies, 1976, Austin: University of Texas
  Press.

\bibitem[{{Deul} \& {den Hartog}(1990)}]{Deul90}
{Deul}, E.~R. \& {den Hartog}, R.~H. 1990, \aap, 229, 362

\bibitem[{{Emerson} \& {Gr\"ave}(1988)}]{Emerson88}
{Emerson}, D.~T. \& {Gr\"ave}, R. 1988, \aap, 190, 353

\bibitem[{{Emerson} {et~al.}(1979){Emerson}, {Klein}, \& {Haslam}}]{Emerson79}
{Emerson}, D.~T., {Klein}, U., \& {Haslam}, C.~G.~T. 1979, \aap, 76, 92

\bibitem[{{Fernini} {et~al.}(1997){Fernini}, {Burns}, \& {Perley}}]{Fernini97}
{Fernini}, I., {Burns}, J.~O., \& {Perley}, R.~A. 1997, \aj, 114, 2292

\bibitem[{{Fletcher} {et~al.}(2011){Fletcher}, {Beck}, {Shukurov},
  {Berkhuijsen}, \& {Horellou}}]{Fletcher11}
{Fletcher}, A., {Beck}, R., {Shukurov}, A., {Berkhuijsen}, E.~M., \&
  {Horellou}, C. 2011, \mnras, 412, 2396

\bibitem[{{Frick} {et~al.}(2016){Frick}, {Stepanov}, {Beck}, {Sokoloff},
  {Shukurov}, {Ehle}, \& {Lundgren}}]{Frick16}
{Frick}, P., {Stepanov}, R., {Beck}, R., {et~al.} 2016, \aap, 585, A21

\bibitem[{{Gr\"ave} {et~al.}(1990){Gr\"ave}, {Klein}, \&
  {Wielebinski}}]{Graeve90}
{Gr\"ave}, R., {Klein}, U., \& {Wielebinski}, R. 1990, \aap, 238, 39

\bibitem[{{Haslam}(1974)}]{Haslam74}
{Haslam}, C.~G.~T. 1974, \aaps, 15, 333

\bibitem[{{Heesen} {et~al.}(2014){Heesen}, {Brinks}, {Leroy}, {Heald}, {Braun},
  {Bigiel}, \& {Beck}}]{Heesen14}
{Heesen}, V., {Brinks}, E., {Leroy}, A.~K., {et~al.} 2014, \aj, 147, 103

\bibitem[{{Helfer} {et~al.}(2003){Helfer}, {Thornley}, {Regan}, {Wong},
  {Sheth}, {Vogel}, {Blitz}, \& {Bock}}]{Helfer03}
{Helfer}, T.~T., {Thornley}, M.~D., {Regan}, M.~W., {et~al.} 2003, \apjs, 145,
  259

\bibitem[{{Hoopes} {et~al.}(2001){Hoopes}, {Walterbos}, \& {Bothun}}]{Hoopes01}
{Hoopes}, C.~G., {Walterbos}, R.~A.~M., \& {Bothun}, G.~D. 2001, \apj, 559, 878

\bibitem[{{Horellou} {et~al.}(1992){Horellou}, {Beck}, {Berkhuijsen}, {Krause},
  \& {Klein}}]{Horellou92}
{Horellou}, C., {Beck}, R., {Berkhuijsen}, E.~M., {Krause}, M., \& {Klein}, U.
  1992, \aap, 265, 417

\bibitem[{{Israel} {et~al.}(1975){Israel}, {Goss}, \& {Allen}}]{Israel75}
{Israel}, F.~P., {Goss}, W.~M., \& {Allen}, R.~J. 1975, \aap, 40, 421

\bibitem[{{Jarrett} {et~al.}(2013){Jarrett}, {Masci}, {Tsai}, {Petty},
  {Cluver}, {Assef}, {Benford}, {Blain}, {Bridge}, {Donoso}, {Eisenhardt},
  {Koribalski}, {Lake}, {Neill}, {Seibert}, {Sheth}, {Stanford}, \&
  {Wright}}]{Jarrett13}
{Jarrett}, T.~H., {Masci}, F., {Tsai}, C.~W., {et~al.} 2013, \aj, 145, 6

\bibitem[{{Jog} \& {Combes}(2009)}]{Jog09}
{Jog}, C.~J. \& {Combes}, F. 2009, \physrep, 471, 75

\bibitem[{{Jurusik} {et~al.}(2014){Jurusik}, {Drzazga}, {Jableka}, {Chy{\.z}y},
  {Beck}, {Klein}, \& {We{\.z}gowiec}}]{Jurusik14}
{Jurusik}, W., {Drzazga}, R.~T., {Jableka}, M., {et~al.} 2014, \aap, 567, A134

\bibitem[{{Kamphuis}(1993)}]{Kamphuis93}
{Kamphuis}, J.~J. 1993, PhD thesis, University of Groningen, (1993)

\bibitem[{{Karachentsev} \& {Kudrya}(2014)}]{Karachentsev14}
{Karachentsev}, I.~D. \& {Kudrya}, Y.~N. 2014, \aj, 148, 50

\bibitem[{{Kelson} {et~al.}(1996){Kelson}, {Illingworth}, {Freedman}, {Graham},
  {Hill}, {Madore}, {Saha}, {Stetson}, {Kennicutt}, {Mould}, {Hughes},
  {Ferrarese}, {Phelps}, {Turner}, {Cook}, {Ford}, {Hoessel}, \&
  {Huchra}}]{Kelson96}
{Kelson}, D.~D., {Illingworth}, G.~D., {Freedman}, W.~F., {et~al.} 1996, \apj,
  463, 26

\bibitem[{{Kenney} {et~al.}(1991){Kenney}, {Scoville}, \& {Wilson}}]{Kenney91}
{Kenney}, J.~D.~P., {Scoville}, N.~Z., \& {Wilson}, C.~D. 1991, \apj, 366, 432

\bibitem[{{Klein} {et~al.}(1984){Klein}, {Wielebinski}, \& {Beck}}]{Klein84}
{Klein}, U., {Wielebinski}, R., \& {Beck}, R. 1984, \aap, 135, 213

\bibitem[{{Kuntz} {et~al.}(2003){Kuntz}, {Snowden}, {Pence}, \&
  {Mukai}}]{Kuntz03}
{Kuntz}, K.~D., {Snowden}, S.~L., {Pence}, W.~D., \& {Mukai}, K. 2003, \apj,
  588, 264

\bibitem[{{Lee} \& {Jang}(2012)}]{Lee12}
{Lee}, M.~G. \& {Jang}, I.~S. 2012, \apjl, 760, L14

\bibitem[{{Leroy} {et~al.}(2012){Leroy}, {Bigiel}, {de Blok}, {Boissier},
  {Bolatto}, {Brinks}, {Madore}, {Munoz-Mateos}, {Murphy}, {Sandstrom},
  {Schruba}, \& {Walter}}]{Leroy12}
{Leroy}, A.~K., {Bigiel}, F., {de Blok}, W.~J.~G., {et~al.} 2012, \aj, 144, 3

\bibitem[{{Mao} {et~al.}(2015){Mao}, {Zweibel}, {Fletcher}, {Ott}, \&
  \~{Tabatabaei}}]{Mao15}
{Mao}, S.~A., {Zweibel}, E., {Fletcher}, A., {Ott}, J., \& \~{Tabatabaei}, F.
  2015, \apj, 800, 92

\bibitem[{{Mihos} {et~al.}(2013){Mihos}, {Harding}, {Spengler}, {Rudick}, \&
  {Feldmeier}}]{Mihos13}
{Mihos}, J.~C., {Harding}, P., {Spengler}, C.~E., {Rudick}, C.~S., \&
  {Feldmeier}, J.~J. 2013, \apj, 762, 82

\bibitem[{{Mihos} {et~al.}(2012){Mihos}, {Keating}, {Holley-Bockelmann},
  {Pisano}, \& {Kassim}}]{Mihos12}
{Mihos}, J.~C., {Keating}, K.~M., {Holley-Bockelmann}, K., {Pisano}, D.~J., \&
  {Kassim}, N.~E. 2012, \apj, 761, 186

\bibitem[{{Mulcahy} {et~al.}(2014){Mulcahy}, {Horneffer}, {Beck}, {Heald},
  {Fletcher}, {Scaife}, {Adebahr}, {Anderson}, {Bonafede}, {Br{\"u}ggen},
  {Brunetti}, {Chy{\.z}y}, {Conway}, {Dettmar}, {En{\ss}lin}, {Haverkorn},
  {Horellou}, {Iacobelli}, {Israel}, {Junklewitz}, {Jurusik}, {K{\"o}hler},
  {Kuniyoshi}, {Orr{\'u}}, {Paladino}, {Pizzo}, {Reich}, \&
  {R{\"o}ttgering}}]{Mulcahy14}
{Mulcahy}, D.~D., {Horneffer}, A., {Beck}, R., {et~al.} 2014, \aap, 568, A74

\bibitem[{{Ohno} \& {Shibata}(1993)}]{Ohno93}
{Ohno}, H. \& {Shibata}, S. 1993, \mnras, 262, 953

\bibitem[{{Oppermann} {et~al.}(2012){Oppermann}, {Junklewitz}, {Robbers},
  {Bell}, {En{\ss}lin}, {Bonafede}, {Braun}, {Brown}, {Clarke}, {Feain},
  {Gaensler}, {Hammond}, {Harvey-Smith}, {Heald}, {Johnston-Hollitt}, {Klein},
  {Kronberg}, {Mao}, {McClure-Griffiths}, {O'Sullivan}, {Pratley}, {Robishaw},
  {Roy}, {Schnitzeler}, {Sotomayor-Beltran}, {Stevens}, {Stil}, {Sunstrum},
  {Tanna}, {Taylor}, \& {Van Eck}}]{Oppermann12}
{Oppermann}, N., {Junklewitz}, H., {Robbers}, G., {et~al.} 2012, \aap, 542, A93

\bibitem[{{Ott} {et~al.}(1994){Ott}, {Witzel}, {Quirrenbach}, {Krichbaum},
  {Standke}, {Schalinski}, \& {Hummel}}]{Ott94}
{Ott}, M., {Witzel}, A., {Quirrenbach}, A., {et~al.} 1994, \aap, 284, 331

\bibitem[{{Sandage}(1961)}]{Sandage61}
{Sandage}, A. 1961, in The Hubble atlas of galaxies, 1961, Washington D.C.:
  Carnegie Institution of Washington.

\bibitem[{{Scowen} {et~al.}(1992){Scowen}, {Dufour}, \& {Hester}}]{Scowen92}
{Scowen}, P.~A., {Dufour}, R.~J., \& {Hester}, J.~J. 1992, \aj, 104, 92

\bibitem[{{Soida} {et~al.}(1996){Soida}, {Urbanik}, \& {Beck}}]{Soida96}
{Soida}, M., {Urbanik}, M., \& {Beck}, R. 1996, \aap, 312, 409

\bibitem[{{Sokoloff} {et~al.}(1998){Sokoloff}, {Bykov}, {Shukurov},
  {Berkhuijsen}, {Beck}, \& {Poezd}}]{Sokoloff98}
{Sokoloff}, D.~D., {Bykov}, A.~A., {Shukurov}, A., {et~al.} 1998, \mnras, 299,
  189

\bibitem[{{Suzuki} {et~al.}(2010){Suzuki}, {Kaneda}, {Onaka}, {Nakagawa}, \&
  {Shibai}}]{Suzuki10}
{Suzuki}, T., {Kaneda}, H., {Onaka}, T., {Nakagawa}, T., \& {Shibai}, H. 2010,
  \aap, 521, A48

\bibitem[{{Tabatabaei} {et~al.}(2007{\natexlab{a}}){Tabatabaei}, {Beck},
  {Kr{\"u}gel}, {Krause}, {Berkhuijsen}, {Gordon}, \& {Menten}}]{Tabatabaei07b}
{Tabatabaei}, F.~S., {Beck}, R., {Kr{\"u}gel}, E., {et~al.} 2007{\natexlab{a}},
  \aap, 475, 133

\bibitem[{{Tabatabaei} {et~al.}(2013{\natexlab{a}}){Tabatabaei}, {Berkhuijsen},
  {Frick}, {Beck}, \& {Schinnerer}}]{Tabatabaei13a}
{Tabatabaei}, F.~S., {Berkhuijsen}, E.~M., {Frick}, P., {Beck}, R., \&
  {Schinnerer}, E. 2013{\natexlab{a}}, \aap, 557, A129

\bibitem[{{Tabatabaei} {et~al.}(2007{\natexlab{b}}){Tabatabaei}, {Krause}, \&
  {Beck}}]{Tabatabaei07c}
{Tabatabaei}, F.~S., {Krause}, M., \& {Beck}, R. 2007{\natexlab{b}}, \aap, 472,
  785

\bibitem[{{Tabatabaei} {et~al.}(2013{\natexlab{b}}){Tabatabaei}, {Schinnerer},
  {Murphy}, {Beck}, {Groves}, {Meidt}, {Krause}, {Rix}, {Sandstrom}, {Crocker},
  {Galametz}, {Helou}, {Wilson}, {Kennicutt}, {Calzetti}, {Draine}, {Aniano},
  {Dale}, {Dumas}, {Engelbracht}, {Gordon}, {Hinz}, {Kreckel}, {Montiel}, \&
  {Roussel}}]{Tabatabaei13b}
{Tabatabaei}, F.~S., {Schinnerer}, E., {Murphy}, E.~J., {et~al.}
  2013{\natexlab{b}}, \aap, 552, A19

\bibitem[{{Urbanik} {et~al.}(1997){Urbanik}, {Elstner}, \& {Beck}}]{Urbanik97}
{Urbanik}, M., {Elstner}, D., \& {Beck}, R. 1997, \aap, 326, 465

\bibitem[{{van Dokkum} {et~al.}(2014){van Dokkum}, {Abraham}, \&
  {Merritt}}]{vanDokkum14}
{van Dokkum}, P.~G., {Abraham}, R., \& {Merritt}, A. 2014, \apjl, 782, L24

\bibitem[{{Van Eck} {et~al.}(2015){Van Eck}, {Brown}, {Shukurov}, \&
  {Fletcher}}]{vanEck15}
{Van Eck}, C.~L., {Brown}, J.~C., {Shukurov}, A., \& {Fletcher}, A. 2015, \apj,
  799, 35

\bibitem[{{Waller} {et~al.}(1997){Waller}, {Bohlin}, {Cornett}, {Fanelli},
  {Freedman}, {Hill}, {Madore}, {Neff}, {Offenberg}, {O'Connell}, {Roberts},
  {Smith}, \& {Stecher}}]{Waller97}
{Waller}, W.~H., {Bohlin}, R.~C., {Cornett}, R.~H., {et~al.} 1997, \apj, 481,
  169

\bibitem[{{Walter} {et~al.}(2008){Walter}, {Brinks}, {de Blok}, {Bigiel},
  {Kennicutt}, {Thornley}, \& {Leroy}}]{Walter08}
{Walter}, F., {Brinks}, E., {de Blok}, W.~J.~G., {et~al.} 2008, \aj, 136, 2563

\bibitem[{{Wardle} \& {Kronberg}(1974)}]{Wardle74}
{Wardle}, J.~F.~C. \& {Kronberg}, P.~P. 1974, \apj, 194, 249

\bibitem[{{Zasov} \& {Abramova}(2006)}]{Zasov06}
{Zasov}, A.~V. \& {Abramova}, O.~V. 2006, Astronomy Reports, 50, 874

\end{thebibliography}

\end{document}